\documentclass[12pt]{article}
 
\usepackage{amsmath}
\usepackage{amssymb}
\usepackage{bm}
\usepackage[pdftex]{graphicx}
\usepackage{bmpsize}
\usepackage{cite}

\newcommand{\sectiono}[1]{\section{#1}\setcounter{equation}{0}}

\newcommand{\bd}[1]{\boldsymbol{#1}}
\newcommand{\id}{\mathbb{I}}

\newcommand{\idSH}{\mathbb{I}_{\mathcal{SH}}}

\begin{document}

\baselineskip=17pt

\begin{titlepage}
%\rightline{\tt arXiv:****.*****}
\rightline{\tt YITP-19-08}
%\rightline\today
\begin{center}
\vskip 2.5cm
{\Large \bf {Heterotic string field theory
 with cyclic $L_\infty$ structure}}
\vskip 1.0cm
{\large {Hiroshi Kunitomo${}^1$ and Tatsuya Sugimoto${}^2$}}
\vskip 1.0cm
{\it {Center for Gravitational Physics}}, 
{\it {Yukawa Institute for Theoretical Physics}}\\
{\it {Kyoto University}},
{\it {Kyoto 606-8502, Japan}}\\
${}^1$kunitomo@yukawa.kyoto-u.ac.jp,\ ${}^2$tatsuya.sugimoto@yukawa.kyoto-u.ac.jp

\vskip 2.0cm

{\bf Abstract}
\end{center}

\noindent
We construct a complete heterotic string field theory that includes both
the Neveu-Schwarz and Ramond sectors. 
We give a construction of general string products, which
realizes a cyclic $L_\infty$ structure and thus provides with a gauge-invariant 
action in the homotopy algebraic formulation. 
Through a map of the string fields, we also give the Wess-Zumino-Witten-like 
action in the large Hilbert space, and verify its gauge invariance 
independently.

\end{titlepage}

\tableofcontents

\newpage

\sectiono{Introduction}
\label{Intro}

There are three main formulations of superstring field theories:
the formulation based on a homotopy algebraic structure in the small Hilbert
space \cite{Erler:2013xta,Erler:2014eba}, 
the Wess-Zumino-Witten(WZW)-like formulation in the large Hilbert space 
\cite{Berkovits:1995ab,Berkovits:2004xh},
and Sen's formulation with an extra free string field 
\cite{Sen:2015uaa,Konopka:2016grr},
each of which has both advantages and disadvantages.
They are complementary and worth studying independently.
In this paper we focus on the former two formulations since
they are not yet fully established for all the superstring field theories, 
while Sen's formulation is. 

In these formulations important progress has recently been made: 
a complete gauge-invariant action for the open superstring field theory including both 
the Neveu-Schwarz (NS) and Ramond (R) sectors was constructed first 
in the WZW-like formulation \cite{Kunitomo:2015usa} and soon afterwards in 
the homotopy algebraic formulation based on the  $A_\infty$ 
structure \cite{Erler:2016ybs}. 
In spite of these successes and several related 
developments \cite{Erler:2015rra,Erler:2015lya,Erler:2017onq,Matsunaga:2015kra,
Goto:2015pqv,Goto:2016ckh,Erler:2016rxg,Kunitomo:2016kwh}, 
these two formulations are not yet fully satisfactory. 
Gauge-invariant actions are only constructed for the NS sector 
in the heterotic string field theory and for the NS-NS sector 
in the type II superstring field theory. 
The purpose of this paper is to fill in some of these missing pieces
by constructing a complete gauge-invariant action for the heterotic 
string field theory in both the homotopy algebraic and the WZW-like
formulations. 

In string field theory interaction of strings is described by means of
string products. The way to construct their basic part, which defines 
how strings connect, is well established
in bosonic string field theory \cite{Saadi:1989tb,Kugo:1989aa,Zwiebach:1992ie}. 
The problem in superstring field
theories is to find a general prescription to insert appropriate operators
for saturating the picture number and to construct the proper string products 
required for the gauge-invariant action.

In the homotopy algebraic formulation, such a prescription is given 
by taking homotopy algebraic structures, an $A_\infty$ structure for 
the open superstring and an $L_\infty$ structure for the heterotic and 
type II superstring, as a guiding principle. 
Proper string products for the NS sector of the open  and 
heterotic string and for the NS-NS sector of the type II superstring 
have been constructed 
%by introducing another string products, called gauge products,
as a solution of differential equations for its generating function 
\cite{Erler:2013xta,Erler:2014eba}.
Although this was successfully extended to the structure including the Ramond 
sector in the open superstring \cite{Erler:2016ybs}, similar extension
in the heterotic and type II superstring is not sufficient to construct 
gauge-invariant actions because of the lack of cyclicity \cite{Erler:2015lya}.
In this paper we propose a similar but slightly different prescription
to construct string products realizing a \textit{cyclic} $L_\infty$ algebra, and 
construct a complete gauge-invariant action for the heterotic string 
field theory in the homotopy algebraic formulation. 
Then, after confirming that it reproduce the first-quantized four-point amplitudes, 
we rewrite the action in the WZW-like form and also 
construct a complete WZW-like action through a field redefinition.

The paper is organized as follows.
In section \ref{HSFT} we review the homotopy algebraic formulation
for the heterotic string field theory. 
A gauge-invariant action and gauge transformation are constructed
on the assumption that the proper string products realizing a cyclic 
$L_\infty$ structure are given for any combination of the NS and Ramond 
string fields. 
Section \ref{cyclic products} is devoted to a concrete construction of such proper string
products realizing a cyclic $L_\infty$ structure in two steps. 
First we consider an $L_\infty$ algebra, which we call a combined
$L_\infty$ algebra in this paper, respecting
(not the Ramond number but) the cyclic Ramond number. It naturally
decomposes two $L_\infty$ algebras, which can be called the dynamical 
and constraint $L_\infty$ algebras \cite{Erler:2017onq}.
Although these two are neither cyclic nor closed
in the small Hilbert space, the combined $L_\infty$ algebra can be cyclic 
with respect to the simple symmetric symplectic form. 
We give in the second step a similarity transformation which 
transforms two decomposed algebras to a desired cyclic $L_\infty$ 
algebra and the constraint restricting the products in the small Hilbert space.
Then, a concrete prescription to construct the proper string products 
realizing the combined $L_\infty$ algebra is given.
After decomposing the commutator of coderivations into two 
operations projecting onto the definite cyclic Ramond number, 
we propose equations for the generating function of (slightly generalized) 
string products generalizing the $L_\infty$ relation and the closedness
condition in the small Hilbert space. 
Introducing another kind of string products, gauge products, 
we show that the solution of these equations
can be obtained by solving some differential equations of generating 
functions of string and gauge products iteratively. 
We confirm, in section \ref{amplitudes},
that the heterotic string field theory we constructed
reproduces the well-known first-quantized four-point amplitudes
including those with the Ramond external states.
In section \ref{action in WZW} we give
a field redefinition which maps the string fields to those in the WZW-like
formulation.
%large Hilbert space formulation. 
After rewriting the action in the WZW-like
form, we can obtain a complete WZW-like action through this field
redefinition. The gauge invariance of the WZW-like action is 
verified independently without referring to the $L_\infty$ structure. 
After a summary and discussion in section \ref{summary}, 
we add four appendices for details which could not be included in the text.
For the reader unfamiliar with the coalgebraic representation of
$L_\infty$ algebra, we briefly summarize it in Appendix \ref{coalgebra}.
Appendix \ref{counting R} is devoted to discussing the ways to distinguish string
products according to the number of Ramond states.
In Appendix \ref{cyclicity proof} we prove the cyclicity of the string products 
constructed in the text. The identity used in section \ref{action in WZW} is proved 
in Appendix \ref{id cyclic proof}.

\sectiono{Heterotic string field theory
%in small Hilbert space}
in the homotopy algebraic formulation}
\label{HSFT}

Let us first summarize several basics of heterotic string
field theory in the homotopy algebraic formulation. 
The heterotic string field $\Phi$ has two components:
\begin{equation}
 \Phi\ =\ \Phi_{NS} + \Phi_R\ \in\ 
\mathcal{H}\ =\ \mathcal{H}_{NS} + \mathcal{H}_R\,.
\end{equation}
The first component $\Phi_{NS}$ is a Grassmann even NS string field
in the small Hilbert space $\mathcal{H}_{NS}$
at ghost number 2 and picture number $-1$. 
The second component $\Phi_R$ is a Grassmann even
R string field in the small Hilbert space $\mathcal{H}_R$
at ghost number 2 and picture number $-1/2$.
Since the heterotic string is a closed string, the string field $\Phi$ 
is restricted by the closed string constraints:
\begin{equation}
 b_0^-\Phi\ =\ L_0^-\Phi\ =\ 0\,.
\label{restrict closed}
\end{equation}
Additionally the R string field $\Phi_R$ satisfies the condition
\begin{equation}
 XY\Phi_R\ =\ \Phi_R\,,
\label{restricted}
\end{equation}
where $X$ and $Y$ are defined by
\begin{align}
 X\ =&\ -\delta(\beta_0)G_0+\delta'(\beta_0)b_0\,,\\
 Y\ =&\ -2c_0^+\delta'(\gamma_0)\,.
\end{align}
The operator $X$ is the picture-changing operator acting on 
states with picture number $-1/2$ and 
$Y$ is its inverse in the following sense acting on those with picture number $-3/2$.
They are BPZ even and satisfy
\begin{equation}
 XYX\ =\ X\,,\qquad YXY\ =\ Y\,,\qquad [Q\,,\,X]\ =\ 0\,,
\label{XYX}
\end{equation}
from which we can show that the operator $XY$ in the constraint in Eq.\,(\ref{restricted})
is a projection operator. We call the space of the states restricted by
Eq.\,(\ref{restricted}) the restricted Hilbert space, or sometimes simply the restricted space, 
denoted by $\mathcal{H}^{res}$. Note that the BRST operator is closed 
in the restricted space: $XYQXY=QXY$\,.

We define a symplectic form of the small Hilbert space $\omega_s$ by
\begin{equation}
 \omega_s(\Phi_1\,,\Phi_2)\ =\ (-1)^{|\Phi_1|}\langle\Phi_1\,,\Phi_2\rangle\
=\ (-1)^{|\Phi_1|}\langle \Phi_1|c_0^-|\Phi_2\rangle\,,
\label{symp small}
\end{equation}
where $\langle\Phi_1|$ is the BPZ conjugate state of $|\Phi_1\rangle$\,.
The symbol $|\Phi|$ denotes the Grassmann property of the string field 
$\Phi$\,: $|\Phi|=0\ (1)$ if the string field $\Phi$ is Grassmann even (odd).
The symplectic form $\omega_s$ is graded anti-symmetric:
\begin{equation}
\omega_s(\Phi_1\,,\Phi_2)\ =\ -(-1)^{|\Phi_1||\Phi_2|}\omega_s(\Phi_2\,,\Phi_1)\,.
\end{equation}
The BRST charge satisfies
\begin{equation}
 \omega_s(Q\Phi_1\,,\Phi_2)\ =\ -(-1)^{|\Phi_1|}\omega_s(\Phi_1\,,Q\Phi_2)\,.
\end{equation}
The symplectic form $\Omega$ of the restricted Hilbert space is then defined by
\begin{equation}
 \Omega(\Phi_1\,,\Phi_2)\ =\ %\omega(\Phi_1\,,\mathcal{G}^{-1}\Phi_2)\,,
\omega_s(\Phi_{1 NS}\,, \Phi_{2 NS}) + \omega_s(\Phi_{1 R}\,, Y\Phi_{2 R})\,,
\end{equation} 
for restricted fields $\Phi_1$ and $\Phi_2$\,.
%\begin{align}
% XY\Phi_{1R}\ =\ \Phi_{1R}\,\qquad  XY\Phi_{2R}\ =\ \Phi_{2R}\,.
%\end{align} 
We also define here a symplectic form of the large Hilbert space $\omega_l$ by
\begin{equation}
 \omega_l(\varphi_1\,,\varphi_2)\ 
=\ -(-1)^{|\varphi_1|}\langle\varphi_1\,,\varphi_2\rangle_l\ =\
- (-1)^{|\varphi_1|}{}_l\langle\varphi_1|c_0^-|\varphi_2\rangle_l\,,
\end{equation}
for later use, where $\varphi_1$ and $\varphi_2$ are some string fields in the large
Hilbert space $\mathcal{H}_l$\, : $\varphi_1\,,\varphi_2\in\mathcal{H}_l$\,. 
The $\eta$ satisfies
\begin{equation}
 \omega_l(\eta\varphi_1\,,\varphi_2)\ =\ 
-(-1)^{|\varphi_1|}\omega_l(\varphi_1\,,\eta\varphi_2)\,.
\end{equation}
%If one of the arguments, suppose $\varphi_2=\Phi_2$, 
%is in the small Hilbert space, %$\Phi_2\in\mathcal{H}$\,,
%we can relate it to $\omega_s$ as
%
%If $\Phi_1,\Phi_2\in\mathcal{H}$\,, two symplectic forms
%$\omega_l$ and $\omega_s$ can be related as
%
If one of the arguments, suppose $\varphi_2\equiv\Phi_2$\,, is in the small
Hilbert space, we can relate $\omega_l$ to $\omega_s$ as\footnote{
Here we assume that the BPZ inner products in the large and small Hilbert spaces
are related as
$\langle\xi\Phi_1\,,\Phi_2\rangle_l\ =\ \langle\Phi_1\,,\Phi_2\rangle$\,.
We denote the zero mode $\xi_0$ as $\xi$ in this paper.
}
\begin{equation}
\omega_l(\varphi_1\,,\Phi_2)\ =\ 
 \omega_s(\eta\varphi_1\,,\Phi_2)\,.
\label{omega l to s}
\end{equation}
%or equivalently
%\begin{equation}
%\omega_l(\xi\Phi_1\,,\Phi_2)\ =\ 
%- \omega_s(\Phi_1\,,\Phi_2)\,. 
%\label{two symplectic forms}
%\end{equation}
%
In the large Hilbert space, $X$ can be written as the BRST exact form
$X=\{Q,\Xi\}$ with 
\begin{equation}
 \Xi\ =\ \xi + (\Theta(\beta_0)\eta\xi-\xi)P_{-3/2}
+(\xi\eta\Theta(\beta_0)-\xi_0)P_{-1/2}\,,
\end{equation}
where $P_n$ is the projector onto the states with picture number $n$\,.

The kinetic term of the action is written by using these symplectic forms as
\begin{align}
 S_0\ =&\ \frac{1}{2}\Omega(\Phi\,,Q\Phi)
\nonumber\\
=&\ \frac{1}{2}\omega_s(\Phi_{NS}\,,Q\Phi_{NS})
+ \frac{1}{2}\omega_s(\Phi_R\,,YQ\Phi_R)\,.
\end{align}

The other fundamental ingredients of heterotic string field theory
are multi-(closed)string products,
\begin{equation}
 L_n(\Phi_1,\cdots,\Phi_n)\,,\qquad (n\ge1)\,,
\label{sproduct}
\end{equation}
which make a string field from $n$ string fields $\Phi_1,\cdots,\Phi_n$\,.
They are graded symmetric under interchange of the $n$ string fields.
If each string product carries proper ghost and picture numbers,
as explained in detail in the next section, the heterotic string interaction 
is described by using such string products with $n\ge2$\,.
In addition, since the heterotic string field in this formulation is in the restricted
small Hilbert space, the string products have also to be closed in the restricted 
space: The R component of Eq.\,(\ref{sproduct}) has also to be in the restricted Hilbert space.
The action of heterotic string field theory 
%in the small Hilbert space 
is written as
\begin{equation}
 S\ =\ \sum_{n=0}^\infty\frac{1}{(n+2)!}\,
\Omega(\Phi\,,L_{n+1}(\underbrace{\Phi\,,\cdots\,,\Phi}_{n+1}))\,, 
\label{small action}
\end{equation}
where the one-string product is identified as the BRST charge: $L_1=Q$\,.
This is invariant under the gauge transformation
\begin{equation}
 \delta\Phi\ =\ \sum_{n=0}^\infty\frac{1}{n!}\, 
L_{n+1}(\underbrace{\Phi\,,\cdots\,,\Phi}_n\,,\Lambda)
\end{equation}
if the string products $L_n$ satisfy the $L_\infty$ relations
\begin{align}
\sum_\sigma\sum_{m=1}^n(-1)^{\epsilon{(\sigma)}}
\frac{1}{m!(n-m)!}
L_{n-m+1}(L_m(\Phi_{\sigma(1)},\cdots,\Phi_{\sigma(m)}), 
\Phi_{\sigma(m+1)},\cdots,\Phi_{\sigma(n)})\ =\ 0
\label{L infinity}
\end{align}
and the cyclicity condition
\begin{equation}
 \Omega(\Phi_1\,, L_n(\Phi_2\,,\cdots\,,\Phi_{n+1}))\ =\ 
- (-1)^{|\Phi_1|}\,\Omega(L_n(\Phi_1\,,\cdots\,,\Phi_n)\,, \Phi_{n+1})\,.
\label{cyclicity}
\end{equation}
Here $\sigma$ in Eq.\,(\ref{L infinity})
denotes the permutation from $\{1,\cdots,n\}$
to $\{\sigma(1),\cdots,\sigma(n)\}$ and the factor
$\epsilon(\sigma)$ is the sign factor of permutation of string fields from
$\{\Phi_1,\cdots,\Phi_n\}$ to $\{\Phi_{\sigma(1)},\cdots,\Phi_{\sigma(n)}\}$\,.
If the set of string products satisfies these conditions it is called
a cyclic $L_\infty$ algebra.
The problem of constructing the heterotic string field theory reduces to
the problem of constructing a set of string products realizing a cyclic $L_\infty$
algebra.\footnote{We also call them string products with an $L_\infty$ structure.}
%
%
%%%%%%%%%%%%%%%% added for the revised version %%%%%%%%%%%%%%%%%%%%%%%%%%
However, the asymmetry of the symplectic form $\Omega$ between the NS and the R sectors
complicates the construction of string products cyclic across both sectors
\cite{Erler:2015lya}. In the next section we propose a way to construct them in two steps.
%%%%%%%%%%%%%%%%%%%%%%%%%%%%%%%%%%%%%%%%%%%%%%%%%%%%%%%%%%%%%%%%%%%%%%

\sectiono{ %String products realizing cyclic $L_\infty$ algebra}
Constructing string products with cyclic $L_\infty$ structure}
\label{cyclic products}

Now let us construct a set of string products realizing a cyclic
$L_\infty$ algebra. We use a coalgebraic representation 
which is convenient to discuss an infinite number of
multi-string products collectively. Its basic definitions and properties
are summarized in Appendix \ref{coalgebra} to make the paper self-contained.

\subsection{Prescription
}

Denote a coderivation corresponding to an $n+2$ string product ($n\ge0$) with picture 
number $p\ge0$\, as $\bd{B}^{(p)}_{n+2}$\,. 
As mentioned in the previous section, string products have to have proper
ghost and picture numbers to describe interactions of heterotic strings.
Here we only need to consider the picture number since the heterotic string field 
$\Phi$ has the same ghost number as the bosonic closed string field. 

To describe the heterotic string interaction the output string state has to have 
the same picture number as the heterotic string field: 
the picture number of its NS (R) component has to be equal to $-1$ $(-1/2)$. 
In order to discuss this kind of picture number counting 
it is useful to introduce the Ramond number
and the cyclic Ramond number \cite{Erler:2015lya,Erler:2016rxg} by
\begin{equation}
 \begin{pmatrix}
  \textrm{Ramond}\\
 \textrm{cyclic Ramond}
 \end{pmatrix}
\textrm{number}\ =\ \#\ \textrm{of Ramond inputs}\ \mp\ \#\ \textrm{of Ramond outputs}\,.
\end{equation}
Let us first suppose that $2r$ of $n+2$ inputs are the R states.
Since the R (NS) states represent the space-time fermions (bosons), the output 
is an NS state. Such a string product is characterized by the Ramond number 
$2r$ and the cyclic Ramond number $2r$\,.
From picture number conservation we have 
\begin{subequations}\label{picture}
\begin{equation}
\left(- \frac{1}{2}\right)\times 2r + (-1)\times(n+2-2r) + p\ =\ -1\,.
\label{pictureNS}
\end{equation}
If $2r+1$ of the inputs are the R states, the output is the R state and 
we have
\begin{equation}
 \left(- \frac{1}{2}\right)\times (2r+1) + (-1)\times(n+1-2r) + p\ =\ -\frac{1}{2}\,.
\label{pictureR}
\end{equation}
\end{subequations}
This is the case characterized by the Ramond number $2r$ and the cyclic Ramond
number $2r+2$\,.
Both of these equations (\ref{picture}) can be solved as %$p=n-r+1$\,, or equivalently
$n=p+r-1$\,.
A candidate coderivation is therefore
\begin{equation}
\bd{Q}+\sum_{p,r=0}^\infty\left(\bd{B}^{(p)}_{p+r+1}|_{2r}^{2r}
+ \bd{B}^{(p)}_{p+r+1}|_{2r}^{2r+2}\right)\
=\ \bd{Q}+\sum_{p,r=0}^\infty\bd{B}^{(p)}_{p+r+1}|_{2r}\,,
\end{equation}
with $\bd{B}^{(0)}_{1}|_0\equiv 0$\,.
However, the cyclicity cannot be transparent in this form
since the Ramond number is not invariant under the ``cyclic permutation'' 
as in Eq.\,(\ref{cyclicity}). 
%%%%%%%%%%%%% changed for the revised version %%%%%%%%%%%%%%%%%%
Instead we consider the string products\footnote{
Here and hereafter we use the convention that a quantity with the Ramond 
or cyclic Ramond number outside the range given in Appendix \ref{counting R} 
is identically equal to zero.}
\begin{equation}
\bd{B}\ \equiv\
 \sum_{p,r=0}^\infty\bd{B}^{(p)}_{p+r+1}|^{2r}\ =\
 \sum_{p,r=0}^\infty\left(\bd{B}^{(p)}_{p+r+1}|^{2r}_{2r}
+ \bd{B}^{(p)}_{p+r+1}|^{2r}_{2r-2}\right)\,,
\label{coderivation B}
\end{equation}
respecting the cyclic Ramond number which is invariant
under the permutation, which makes it possible to construct
the cyclic coderivation,
at the cost, however, that the string products of Eq.\,(\ref{coderivation B})
do not satisfy the condition in Eq.\,(\ref{pictureR}).
%%%%%%%%%%%%%%%%%%%%%%%%%%%%%%%%%%%%%%%%%%%%%%%%%%%%%%%%%%%%%%%%
The picture number deficit of the string products in the second term
is equal to $1$\,: the output of the second term is the R state with 
picture number $-1/2-1=-3/2$\,.
This combination of string products appears naturally as the difference
of two coderivations $\bd{D}$ and $\bd{C}$ with picture number deficit
$0$ and $1$ respectively: 
\begin{equation}
 \bd{D}-\bd{C}\ =\ \bd{Q} - \bd{\eta} + \bd{B}\,,
\end{equation}
with
\begin{align}
 &\pi_1\bd{D}\ =\ 
\pi_1\bd{Q}+\sum_{p,r=0}^\infty\pi_1\bd{B}^{(p)}_{p+r+1}|^{2r}_{2r}\ =\
\pi_1\bd{Q}+\pi_1^0\bd{B}\,,
\label{coder D}\\
 &\pi_1\bd{C}\ =\ 
\pi_1\bd{\eta}-\sum_{p,r=0}^\infty\pi_1\bd{B}^{(p)}_{p+r+1}|^{2r}_{2r-2}\ =\
\pi_1\bd{\eta} - \pi_1^1\bd{B}\,.
\label{coder C}
\end{align}
Here, $\pi_1^0$ ($\pi_1^1$) is the projection operator onto $\mathcal{H}_{NS}$
($\mathcal{H}_R$) introduced in Appendix \ref{coalgebra}.
%%%%%%%%%%%%% add for revised version %%%%%%%%%%%%%
This combination of string products can be cyclic with respect to the symmetric
symplectic form $\omega_s$\,.

Note that if we suppose that 
the coderivation $\bd{D}-\bd{C}$ satisfies the $L_\infty$ relation
\begin{equation}
[\bd{D}-\bd{C},\bd{D}-\bd{C}]=0\,,
\label{D-C}
\end{equation}
it is not closed in the small Hilbert space since $[\bd{\eta},\bd{D}-\bd{C}]\ne0$\,.
As the first step, we construct an $L_\infty$ algebra $\bd{D}-\bd{C}$ cyclic
with respect to the symplectic form $\omega_l$\,, which is comparatively 
easy due to the symmetry of $\omega_l$ between the NS and R sectors.
We denote this cyclic $L_\infty$ algebra as $(\mathcal{H}_l, \omega_l, \bd{D}-\bd{C})$\,.
%%%%%%%%%%%%%%%%%%%%%%%%%%%%%%%%%%%%%%%%%%%%%%%%%%%
%

%%%%%%%%%%  changed for revised version %%%%%%%%%%%%
By decomposing Eq.\,(\ref{D-C}) with the picture number deficit, we can find that 
%the $L_\infty$ relation
%\begin{equation}
%[\bd{D}-\bd{C},\bd{D}-\bd{C}]=0 
%\end{equation}
it is equivalent to the set of conditions
\begin{equation}
[\bd{D}\,,\bd{D}]\ =\ [\bd{C}\,,\bd{C}]\ =\ [\bd{D}\,,\bd{C}]\ =\ 0\,.
\end{equation}
These two mutually commutative $L_\infty$ algebras $\bd{D}$ and $\bd{C}$ are
the heterotic string analogs of the dynamical and constraint $L_\infty$ algebras in
Ref.\cite{Erler:2017onq}, respectively.
They are neither cyclic nor closed in the small Hilbert space.
We denote them as $(\mathcal{H}_l, \bd{D})$\,, $(\mathcal{H}_l, \bd{C})$\,.

Once the cyclic $L_\infty$ algebra
$(\mathcal{H}_l, \omega_l, \bd{D}-\bd{C})$
is constructed, a desired cyclic $L_\infty$ algebra 
$(\mathcal{H}^{res}\,,\Omega\,,\bd{L})$ can be obtained 
in a similar manner to Ref.\cite{Erler:2017onq}.
Let us consider, in the second step, a simultaneous similarity transformation of
$\bd{D}$ and $\bd{C}$ 
generated by an invertible cohomomorphism $\hat{\bd{F}}$\,.
Since a similarity transformation preserves $L_\infty$ structure, 
they are still mutually commutative $L_\infty$ algebras after transformation.
Suppose that $\hat{\bd{F}}$ transforms $\bd{C}$ 
to $\hat{\bd{F}^{-1}}\bd{C}\hat{\bd{F}}=\bd{\eta}$\,. 
Then $\bd{L}\equiv\hat{\bd{F}}^{-1}\bd{D}\hat{\bd{F}}$ is commutative to 
$\bd{\eta}$ and hence an $L_\infty$ algebra in the small Hilbert space
$(\mathcal{H},\bd{L})$\,.
In order that $\bd{L}$ is further an $L_\infty$ algebra in the restricted
space $\mathcal{H}^{res}$\,, %$(\mathcal{H}^{res}, \bd{L})$\,,
the Ramond component of the output $\pi_1^1\bd{L}$ must be 
in the restricted Hilbert space. So we assume here that $\bd{L}$ has the form
\begin{equation}
\pi_1\bd{L}\ =\ \pi_1\bd{Q} + \pi_1^0\bd{b} + X\pi_1^1\bd{b}\,,
\label{coder L}
\end{equation}
with $\bd{b}=\sum_{n=2}^\infty\bd{b}_n$\,.
If $\bd{L}$ has this form, we have
\begin{equation}
 \Omega(\Phi_1\,, L_n(\Phi_2\,,\cdots\,,\Phi_{n+1}))\ =\
 \omega_s(\Phi_1\,, b_n(\Phi_2\,,\cdots\,,\Phi_{n+1}))\,,
\qquad (n\ge2)\,,
\end{equation}
for $\Phi_1\,,\cdots\,,\Phi_{n+1}\in\mathcal{H}^{res}$\,,
and hence the cyclicity of $\bd{L}$ with respect to $\Omega$ 
is translated to simpler cyclicity of $\bd{b}$ with respect to $\omega_s$.
We need two steps since $\bd{b}$ itself cannot be string products 
of an $L_\infty$ algebra since it does not have a definite picture number deficit.

Next, let us show that the desired cohomomorphism $\hat{\bd{F}}$ is concretely given by
\begin{equation}
\pi_1 \hat{\bd{F}}^{-1}\ =\ \pi_1\idSH-\Xi\pi_1^1\bd{B}\,.
\label{F inverse}
\end{equation}
Acting $\hat{\bd{F}}$ from the right of both sides, we have
\begin{equation}
 \pi_1\hat{\bd{F}}\ =\ \pi_1\,\idSH+\Xi\pi^1_1\bd{B}\hat{\bd{F}}\,.
\label{defF}
\end{equation}
%which gives rise to an expected cyclic $L_\infty$ algebra, 
%$(\mathcal{H}^{res},\Omega,\bd{L})$\,.
By decomposing (\ref{F inverse}) we have
\begin{align}
 \pi_1^0\hat{\bd{F}}^{-1}\ =&\ \pi_1^0\idSH\,,\\
 \pi_1^1\hat{\bd{F}}^{-1}\ =&\ \pi_1^1(\idSH-\Xi\bd{B})\ =\ %(\bd{\eta}-\bd{C})\ =\ 
\pi_1^1(\eta \Xi\,\idSH+\Xi\bd{C})\,.
\end{align}
Using $\pi_1^0\bd{C}=\pi_1^0\bd{\eta}$\,,
$\pi_1^0\hat{\bd{F}}=\pi_1^0\hat{\bd{F}}^{-1}\hat{\bd{F}}=\pi_1^0\idSH$ and $\bd{C}^2=0$\,,
we find that
\begin{align}
 \pi_1^0\hat{\bd{F}}^{-1}\bd{C}\hat{\bd{F}}\ =&\ \pi_1^0\bd{\eta}\,,\\
 \pi_1^1\hat{\bd{F}}^{-1}\bd{C}\hat{\bd{F}}\ 
=&\ \eta\Xi\pi_1^1(\bd{\eta}-\bd{B})\hat{\bd{F}} 
\nonumber\\
=&\ \pi_1^1\bd{\eta}(\idSH-\Xi\bd{B})\hat{\bd{F}}\ =\ \pi_1^1\bd{\eta}\,,
\end{align}
and hence 
\begin{equation}
 \pi_1\hat{\bd{F}}^{-1}\bd{C}\hat{\bd{F}}\ =\ \pi_1\bd{\eta}\,.
\end{equation}
Similarly from $[\bd{C},\bd{D}]=0$ and $\pi_1^1\bd{D}=\pi_1^1\bd{Q}$\,, we find that
\begin{align}
 \pi^1_1\bd{B}\bd{D}\ =&\ 
\pi^1_1(\bd{\eta}-\bd{C})\bd{D}\
\nonumber\\
=&\ \bd{\eta}\pi^1_1\bd{Q}+\pi^1_1\bd{Q}\bd{C}
=\ -Q\pi^1_1\bd{B}\,,
\end{align}
and then we can show that
\begin{align}
 \pi_1^0\hat{\bd{F}}^{-1}\bd{D}\hat{\bd{F}}\ =&\ 
\pi_1^0(\bd{Q}+\bd{B}\hat{\bd{F}})\,,\\
%%%%%%%%%%%%%%%%%%
 \pi_1^1\hat{\bd{F}}^{-1}\bd{D}\hat{\bd{F}}\ 
=&\ \pi_1^1(\bd{Q} + X\bd{B}\hat{\bd{F}})\,,
\end{align}
which provides an expected form of $\bd{L}$ as
\begin{align}
\pi_1\bd{L}\ =\ \pi_1\hat{\bd{F}}^{-1}\bd{D}\hat{\bd{F}}\ 
=\ \pi_1\bd{Q}+\pi_1^0\bd{B}\hat{\bd{F}}+X\pi_1^1\bd{B}\hat{\bd{F}}\,.
\label{L hetero}
\end{align}

%%%%%%%%%%%%  changed for revised version %%%%%%%%%%%%%%%%
Finally, we can also show, as proved in Appendix \ref{cyclicity proof}, 
that if $\bd{B}$ is cyclic with respect to $\omega_l$,
then $\pi_1\bd{b}=\pi_1\bd{B}\hat{\bd{F}}$ is cyclic with respect to $\omega_s$\,.
%%%%%%%%%%%%%%%%%%%%%%%%%%%%%%%%%%%%%%%%%%%%%%%%%%%%%%%%%%
In this way we can obtain the desired cyclic $L_\infty$ algebra 
$(\mathcal{H}^{res}, \Omega, \bd{L})$
from a cyclic $L_\infty$ algebra $(\mathcal{H}_l,\omega_l,\bd{D}-\bd{C})$\,.

\subsection{Explicit construction
}

Now, the remaining task is to construct concretely a cyclic $L_\infty$ algebra
$(\mathcal{H}_l,\omega_l,\bd{D}-\bd{C})$\,. 
Let us start with considering the string products with picture number zero
$\bd{L}^{(0)}=\sum_{n=0}^\infty \bd{L}^{(0)}_{n+1}$\,.
The cyclic $L_\infty$ algebra $(\mathcal{H}_l, \omega_l,\bd{L}^{(0)})$
can easily be constructed in the same way as in the bosonic closed string
field theory. Hereafter we call it the bosonic $L_\infty$ algebra, which is
assumed to be known.
We define a generating function 
\begin{equation}
 \bd{L}^{(0)}(s)\ =\  
\bd{Q} + \sum_{m,r=0}^\infty s^m\bd{L}^{(0)}_{m+r+1}|^{2r}\
\equiv\ \bd{Q} + \bd{L}_B^{(0)}(s)\,;
\label{bosonic generating}
\end{equation}
counting the picture number deficit, the picture number of its NS component 
$\pi_1^0\bd{L}^{(0)}$ is $-1-m$\,, and that of its R component 
$\pi_1^1\bd{L}^{(0)}$ is $-3/2-m$\,. It reduces to $\bd{L}^{(0)}$ at $s=1$\,.
From the $L_\infty$ relation $[\bd{L}^{(0)},\bd{L}^{(0)}]=0$\,, we can show that
$\bd{L}^{(0)}_B(s)$ satisfies
\begin{subequations}\label{bosonic L}  
\begin{equation}
 [\,\bd{Q}\,, \bd{L}_B^{(0)}(s)]+\frac{1}{2}[\bd{L}_B^{(0)}(s)\,, \bd{L}_B^{(0)}(s)]^1
+ \frac{s}{2}[\,\bd{L}_B^{(0)}(s)\,, \bd{L}_B^{(0)}(s)]^2\ =\ 0\,,
\label{bosonic L infinity1}
\end{equation}
and is closed in the small Hilbert space
\begin{equation}
 [\,\boldsymbol{\eta}\,, \bd{L}_B^{(0)}(s)]\ =\ 0\,,
\label{bosonic L infinity2}
\end{equation}
\end{subequations}
where $[\ ,\ ]^{1,2}$
are the operations introduced in Eq.\,(\ref{two operations}),
which are obtained by projecting the commutator onto the components
with the definite cyclic Ramond number.

On the other hand, the $L_\infty$ relation $[\boldsymbol{D}-\boldsymbol{C},
\boldsymbol{D}-\boldsymbol{C}]=0$ is equivalent to the equations
\begin{subequations} \label{hetero L} 
\begin{align}
 [\,\boldsymbol{Q}\,, \boldsymbol{B}(t)\,] 
+ \frac{1}{2}[\,\boldsymbol{B}(t)\,, \boldsymbol{B}(t)\,]^1\ =&\ 0\,,
\label{eq B1}\\
 [\,\boldsymbol{\eta}\,, \boldsymbol{B}(t)\,] 
- \frac{t}{2}[\,\boldsymbol{B}(t)\,, \boldsymbol{B}(t)\,]^2\ =&\ 0\,,
\label{eq B2}
\end{align}
\end{subequations}
for the generating function counting the picture number,
\begin{equation}
 \bd{B}(t)\ =\ \sum_{n,r=0}^\infty t^n \bd{B}^{(n)}_{n+r+1}|^{2r}\,.
\label{counting p number}
\end{equation}
We extend Eqs.\,(\ref{hetero L}) to the equations
\begin{subequations} \label{IJst}
 \begin{align}
 \bd{I}(s,t)\ \equiv&\ [\,\bd{Q}\,,\,\bd{B}(s,t)\,]
+\frac{1}{2}[\,\bd{B}(s,t)\,,\,\bd{B}(s,t)\,]^1
+\frac{s}{2}[\,\bd{B}(s,t)\,,\,\bd{B}(s,t)\,]^2\ =\ 0\,,\\
\bd{J}(s,t)\ \equiv&\ [\,\bd{\eta}\,,\,\bd{B}(s,t)\,]
-\frac{t}{2}[\,\bd{B}(s,t)\,,\,\bd{B}(s,t)\,]^2\ =\ 0
\end{align}
\end{subequations}
by introducing the string products with non-zero picture number
deficit counted by $s$ in
\begin{equation}
 \bd{B}(s,t)\ 
=\ \sum_{m,n,r=0}^\infty s^m t^n \bd{B}^{(n)}_{m+n+r+1}|^{2r}\
=\  \sum_{n=0}^\infty t^n\,\bd{B}^{(n)}(s)\,.
\end{equation}
The desired string products are included as $\bd{B}(t)=\bd{B}(0,t)$
since Eqs.\,(\ref{IJst}) reduce to Eqs.\,(\ref{hetero L}) at $s=0$\,.
In the following we construct the string products 
satisfying Eqs.\,(\ref{IJst}).

Let us first show that such string products %$\bd{B}(s,t)$ 
can be obtained by postulating the differential equations
\begin{subequations}\label{diff}
 \begin{align}
 \partial_t\bd{B}(s,t)\ =&\
[\bd{Q}\,,\,\bd{\lambda}(s,t)]
+ [\bd{B}(s,t)\,,\,\bd{\lambda}(s,t)]^1
+ s[\bd{B}(s,t)\,,\,\bd{\lambda}(s,t)]^2\,,
\label{diff 1}
\\
[\bd{\eta}\,, \bd{\lambda}(s,t)]\ =&\  
\partial_s\bd{B}(s,t)
+ t\,[\bd{B}(s,t)\,, \bd{\lambda}(s,t)]^2\,,
\label{diff 2}
\end{align}
\end{subequations}
by introducing a degree-even coderivation
\begin{equation}
\bd{\lambda}(s,t)\ 
=\ \sum_{m,n,r=0}^\infty s^m t^n\bd{\lambda}^{(n+1)}_{m+n+r+2}|^{2r}\
=\ \sum_{n=0}^\infty t^n\,\bd{\lambda}^{(n+1)}(s)\,,
\end{equation} 
which is an analog of the gauge products in 
Ref.\cite{Erler:2014eba}.
Equations (\ref{diff}) imply
\begin{align}
 \partial_t\bd{I}(s,t)\ =&\ [\bd{I}(s,t)\,, \bd{\lambda}(s,t)]^1 
+ s\,[\bd{I}(s,t)\,, \bd{\lambda}(s,t)]^2\,,
\label{del I}\\
%%%%%%%%
 \partial_t\bd{J}(s,t)\ =&\ 
- \partial_s\bd{I}(s,t)
-t[\bd{I}(s,t)\,, \bd{\lambda}(s,t)]^2 
+ [\bd{J}(s,t)\,, \bd{\lambda}(s,t)]^1 + s[\bd{J}(s,t)\,, \bd{\lambda}(s,t)]^2\,.
\label{del J}
\end{align}
Since these equations are homogeneous in $\bd{I}(s,t)$ and $\bd{J}(s,t)$\,,
Eqs.\,(\ref{IJst}) follow from the equations at $t=0$ which agree
with Eqs.\,(\ref{bosonic L}) satisfied by the bosonic products
$\bd{L}^{(0)}_B(s)$ assumed to be known.
All the products $\bd{B}(s,t)$ satisfying Eqs.\,(\ref{IJst}) can be determined by solving the differential
equations in Eqs.\,(\ref{diff}) under the condition 
\begin{equation}
 \bd{B}(s,0)=\bd{B}^{(0)}(s)=\bd{L}^{(0)}_B(s)\,.
\label{B(0)}
\end{equation}

\begin{figure}
 \begin{center}
 \includegraphics{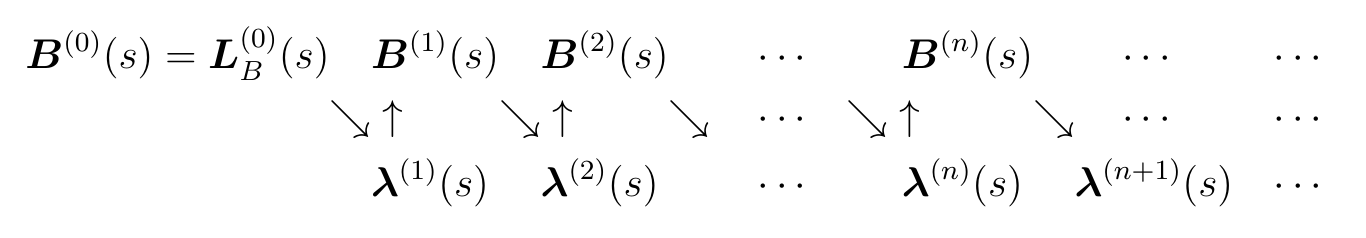}
\caption{The flow along which
all the $\bd{B}^{(n)}(s)$ and $\bd{\lambda}^{(n)}(s)$ are 
determined recursively from 
%the initial condition 
$\bd{B}^{(0)}(s)=\bd{L}^{(0)}_B(s)$.}
\label{how to solve}
 \end{center}
\end{figure}

Finally, let us find a concrete expression
of the string products.
By expanding Eqs.\,(\ref{diff}) in $t$, we obtain
\begin{subequations}\label{eqs p}   
\begin{align}
 (n+1) \bd{B}^{(n+1)}(s)\ =&\ [\bd{Q}\,,\bd{\lambda}^{(n+1)}(s)]
\nonumber\\
&\
+\sum_{n'=0}^n[\bd{B}^{(n-n')}(s)\,,\bd{\lambda}^{(n'+1)}(s)]^1
+\sum_{n'=0}^n s\,[\bd{B}^{(n-n')}(s)\,,\bd{\lambda}^{(n'+1)}(s)]^2\,,
\label{B general}\\
%%%%%%%%%%%%%%%%
[\bd{\eta}\,,\bd{\lambda}^{(n+1)}(s)]\ =&\
\partial_s\bd{B}^{(n)}(s)\
%=&\
%[\bd{\eta}\,,\bd{\lambda}^{(n+1)}(s)]
+
%-
\sum_{n'=0}^{n-1}[\bd{B}^{(n-n'-1)}(s)\,,
\bd{\lambda}^{(n'+1)}(s)]^2\,,
\label{lambda general}
\end{align}
\end{subequations}
which can be solved iteratively.
For $n=0$ we have
%Substituting it into (\ref{diff}) with $t=0$\,, we have
\begin{subequations} 
\begin{align}
\bd{B}^{(1)}(s)\ =&\ 
[\bd{Q}\,, \bd{\lambda}^{(1)}(s)] + [\bd{L}^{(0)}_B(s)\,, \bd{\lambda}^{(1)}(s)]^1
+ s\, [\bd{L}^{(0)}_B(s)\,, \bd{\lambda}^{(1)}(s)]^2\,,
\label{B1}\\
%%%%%
[\bd{\eta}\,, \bd{\lambda}^{(1)}(s)]\ =&\
\partial_s\bd{L}^{(0)}_B(s)\,.
\label{lambda1}
\end{align}
\end{subequations}
Since $[\,\bd{\eta},\bd{L}^{(0)}_B(s)]=0$\,,
we can consistently determine the gauge products $\bd{\lambda}^{(1)}(s)$ 
from Eq.\,(\ref{lambda1}) as
\begin{align}
 \bd{\lambda}^{(1)}(s)\ =&\ \sum_{m,r=0}^\infty (m+1)s^m
\frac{1}{m+r+3}\left(
\xi\,L^{(0)}_{m+r+2}|^{2r} -
L^{(0)}_{m+r+2}|^{2r}(\xi\,\wedge\,\id_{m+r+1})\right)
\nonumber\\
\equiv&\ \xi\circ\partial_s\bd{L}^{(0)}_B(s)\,,
\label{lambda(1)}
\end{align}
respecting the cyclicity.
Then, all the terms on the right-hand side of Eq.\,(\ref{B1}) are given,
which determines $\bd{B}^{(1)}(s)$\,. % as
In order to go to the next step and beyond, we note here that 
the right-hand side of Eq.\,(\ref{lambda general}),
\begin{align}
\bd{K}^{(n)}(s)\ \equiv\
\partial_s\bd{B}^{(n)}(s)
+\sum_{m=0}^{n-1}[\,\bd{B}^{(n-m-1)}(s)\,,\,\bd{\lambda}^{(m+1)}(s)\,]^2\,,
\end{align}
satisfies
\begin{align}
 [\,\bd{\eta}\,,\,\bd{K}^{(n)}(s)\,]\ =&\
\partial_s\bd{J}^{(n)}(s)+\sum_{n'=0}^{n-1}[\,\bd{J}^{(n-n'-1)}(s)\,,\,\bd{\lambda}^{(n'+1)}(s)\,]^2
\nonumber\\
&\
-\sum_{n'=0}^{n-1}[\,\bd{B}^{(n-n'-1)}(s)\,,\,\Big([\,\bd{\eta}\,,
\,\bd{\lambda}^{(n'+1)}(s)\,]-\bd{K}^{(n')}(s)\Big)\,]^2\,,
\label{eta K}
\end{align}
where
\begin{equation}
\bd{J}^{(n)}(s)\ \equiv\ \frac{1}{n!}\partial_t^n\bd{J}(s,t)|_{t=0}\
=\ [\,\bd{\eta}\,,\,\bd{B}^{(n)}(s) \,]
-\frac{1}{2}\sum_{n'=0}^{n-1}[\,\bd{B}^{(n-n'-1)}(s)\,,\,\bd{B}^{(n')}(s) \,]^2\,.
\end{equation}
In the next step, $n=1$\,, this becomes
\begin{align}
 [\bd{\eta}\,\, \bd{K}^{(1)}(s)]\ =&\ \partial_s\bd{J}^{(1)}(s) +
[\bd{J}^{(0)}(s)\,,\,\bd{\lambda}^{(1)}(s)]^2 
- [\bd{B}^{(0)}(s)\,,\,\Big([\bd{\eta}\,,\,\bd{\lambda}^{(1)}(s)]-\bd{K}^{(0)}(s)\Big)]
\nonumber\\
=&\ 0\,,
\end{align}
since $\bd{B}^{(0)}(s)$\,, $\bd{B}^{(1)}(s)$ and
$\bd{\lambda}^{(1)}(s)$ are already determined so as to satisfy 
\begin{equation}
 \bd{J}^{(0)}(s)\ =\ \bd{J}^{(1)}(s)\ =\
  [\bd{\eta}\,,\,\bd{\lambda}^{(1)}(s)]-\bd{K}^{(0)}(s)\ =\ 0\,.
\end{equation}
This enables us to determine $\bd{\lambda}^{(2)}(s)$ consistently from 
Eq.\,(\ref{lambda general}) with $n=1$\,, 
\begin{equation}
[\bd{\eta},\bd{\lambda}^{(2)}]-\bd{K}^{(1)}=0\,,
\end{equation}
as
\begin{align}
 \bd{\lambda}^{(2)}(s)\ =\ \xi\circ\Big(
\partial_s\bd{B}^{(1)}(s) + [\bd{B}^{(0)}(s)\,,\,\bd{\lambda}^{(1)}(s)]^2\Big)\,,
\end{align}
%%  ]^2 2019/10/2
and to determine $\bd{B}^{(2)}$ from Eq.\,(\ref{B general}) with $n=1$\,. 
Similarly, when we solved Eqs.\,(\ref{eqs p}) iteratively
and determined $\bd{\lambda}^{(n)}(s)$ and $\bd{B}^{(n)}(s)$ for $n\le n_0$\,,
the equations
\begin{equation}
 \bd{J}^{(n)}(s)\ =\ 
[\,\bd{\eta}\,,\,\bd{\lambda}^{(n)}(s) \,]-\bd{K}^{(n-1)}(s)\ =\ 0
\end{equation}
hold for $n\le n_0$ by construction,
which provides $[\,\bd{\eta},\bd{K}^{(n_0)}\,]=0$ from Eq.\,(\ref{eta K})
and guarantees to go one step forward consistently.
Thus we can recursively determine 
%$\bd{B}(s,t)$ and $\bd{\lambda}(s,t)$ 
all the $\bd{B}^{(n)}(s)$ and $\bd{\lambda}^{(n)}(s)$\,, hence $\bd{B}(s,t)$ 
and $\bd{\lambda}(s,t)$\,, along the flow depicted in Fig.\ref{how to solve}.
The concrete expression can be obtained through the recursive relations
\begin{subequations} 
 \begin{align}
 \bd{\lambda}^{(n+1)}(s)\ =&\ \xi\circ\Big(
\partial_s\bd{B}^{(n)}(s)\
+\sum_{n'=0}^{n-1}[\bd{B}^{(n-n'-1)}(s)\,,
\bd{\lambda}^{(n'+1)}(s)]^2
\Big)\,,\\
 (n+1) \bd{B}^{(n+1)}(s)\ =&\ [\bd{Q}\,,\bd{\lambda}^{(n+1)}(s)]
\nonumber\\
&\
+\sum_{n'=0}^n[\bd{B}^{(n-n')}(s)\,,\bd{\lambda}^{(n'+1)}(s)]^1
+\sum_{n'=0}^n s\,[\bd{B}^{(n-n')}(s)\,,\bd{\lambda}^{(n'+1)}(s)]^2\,,  
\end{align}
\end{subequations}
starting from the initial condition $\bd{B}^{(0)}(s)=\bd{L}^{(0)}_B(s)$\,.
All the $\bd{B}(s,t)$ and $\bd{\lambda}(s,t)$ are cyclic with respect to
$\omega_l$ by construction. Thus a cyclic $L_\infty$
$(\mathcal{H}_l,\omega_l,\bd{D}-\bd{C})$ is obtained.

\sectiono{Four-point amplitudes}
\label{amplitudes}

It is interesting how the heterotic string field theory we have constructed reproduces
the first-quantized amplitudes. In this section we demonstrate it by focusing on
the four-point amplitudes as the simplest example.
We concretely calculate three types of four-point amplitudes, four-NS\,, two-NS-two-R,
and four-R, following a similar process to the one given in Ref.\cite{Erler:2013xta}. 
In this section we denote the string fields $\Phi_{NS}$ and $\Phi_R$ as $\Phi$ 
and $\Psi$\,, respectively, for notational simplicity.
We take the Siegel gauge satisfying the conditions
\begin{equation} 
b_0^+\Phi\ =\ \beta_0\Psi\ =\ 0\,.
\end{equation}
The propagators in this gauge are obtained in the usual way as
\begin{equation}
 \Pi_{NS}\ 
=\  -\frac{b_0^+b_0^-}{L_0^+}\mathcal{P}\,,
\qquad
 \Pi_R\ =\ -\frac{b_0^+b_0^-X}{L_0^+}\mathcal{P}\,,
\end{equation}
where %$\mathcal{P}$ is the projection operator
\begin{equation}
 \mathcal{P}\ \equiv\ \int^{2\pi}_0\frac{d\theta}{2\pi}e^{i\theta L_0^-}\,.
\end{equation}

\subsection{Four-NS amplitude}

To warm up let us first consider the four-NS amplitude, 
although it can be calculated without using the new results obtained in this paper.
The first-quantized four-NS amplitude is expressed, for example, 
in the form
\begin{equation}
 \mathcal{A}_4^\textrm{1st}(\Phi_1\,,\Phi_2\,,\Phi_3\,,\Phi_4)\ =\ \int d^2z\,
\langle\left((X_0)^2\Phi_1(0)\right)(b_{-1}^+b_{-1}^-\Phi_2(z))\Phi_3(1)
\Phi_4(\infty)\rangle\rangle\,,
\end{equation}
where $X_0=\{Q,\xi\}$ and
$\Phi_1,\cdots,\Phi_4$ are on-shell physical NS vertex operators in the $-1$ picture:
it satisfies $Q\Phi=0$\,.
The correlator is evaluated in the small Hilbert space on the complex $z$-plane.
We put two $X_0$s in front of $\Phi_1$\,, which is 
possible since the physical amplitudes are independent of the position of 
the picture-changing operators.
Since the structure of bosonic moduli is common to the bosonic closed string,
we can express this using the bosonic string products $L_n^{(0)}$ as
\begin{align}
 \mathcal{A}_4^\textrm{1st}(\Phi_1\,,\Phi_2\,,\Phi_3\,,\Phi_4)\ 
=&\ \omega_s\Bigg((X_0)^2\Phi_1\,,\Big(
L_3^{(0)}(\Phi_2\,,\Phi_3\,,\Phi_4)
- L_2^{(0)}(\Phi_2\,,\frac{b_0^+}{L^+_0}L_2^{(0)}(\Phi_3\,,\Phi_4))
\nonumber\\
&\hspace{10mm}
- L_2^{(0)}(\Phi_3\,,\frac{b_0^+}{L^+_0}L_2^{(0)}(\Phi_4\,,\Phi_2))
- L_2^{(0)}(\Phi_4\,,\frac{b_0^+}{L^+_0}L_2^{(0)}(\Phi_2\,,\Phi_3))
\Big)\Bigg)\,.
\label{4NS 1st}
\end{align}
The moduli integral $b_0^-\mathcal{P}$ is hidden behind the definition of 
the linear map $L_n^{(0)}$\,.
This can be regarded as a multi-linear map:
\begin{equation}
 \langle\mathcal{A}_4|\, :\ \mathcal{H}^{NS}_Q\otimes(\mathcal{H}^{NS}_Q)^{\wedge3}\
\longrightarrow\ \mathbb{C}\,,
\end{equation}
where $\mathcal{H}^{NS}_Q\subset\mathcal{H}_{NS}$ is the subspace of states annihilated by
$Q$\,. Putting $\Phi_1,\cdots,\Phi_4$ out, we can express Eq.\,(\ref{4NS 1st}) as
\begin{equation}
 \langle\mathcal{A}_4^{1st}|\ =\ \langle\omega_s|(X_0)^2\otimes
\Big(L_3^{(0)}-L_2^{(0)}\big(\id\wedge\,\frac{b_0^+}{L^+_0}L_2^{(0)}\big)\Big)\,.
\label{A4 1st map}
\end{equation}
Here the $\langle\omega_s|$ is a bilinear map 
representation of the symplectic form $\omega_s$\,.
We can also write Eq.\,(\ref{A4 1st map}) using the coderivations as
\begin{equation}
 \langle\mathcal{A}_4^{1st}|\ =\ \langle\omega_s|(X_0)^2\otimes
\pi_1^0\Big(\bd{L}_3^{(0)}|^0_0-\bd{L}_2^{(0)}|^0_0\,\frac{b_0^+}{L^+_0}\bd{L}_2^{(0)}|^0_0\Big)\,,
\label{4NS 1st coderivation}
\end{equation}
where $\frac{b_0^+}{L_0^+}\bd{L}_2^{(0)}$ is the coderivation derived from 
$\frac{b_0^+}{L_0^+}L_2^{(0)}$\,.

On the other hand, the four-NS amplitude obtained from the heterotic string field theory is given
by
\begin{align}
 \mathcal{A}_4(\Phi_1\,,\Phi_2\,,\Phi_3\,,\Phi_4)\ 
=&\ \omega_s\Bigg(\Phi_1\,,\Big(
L_3^{(2)}(\Phi_2\,,\Phi_3\,,\Phi_4)
- L_2^{(1)}(\Phi_2\,,\frac{b_0^+b_0^-}{L^+_0}\mathcal{P}c_0^-L_2^{(1)}(\Phi_3\,,\Phi_4))
\nonumber\\
&\hspace{8mm}
- L_2^{(1)}(\Phi_3\,,\frac{b_0^+b_0^-}{L^+_0}\mathcal{P}c_0^-L_2^{(1)}(\Phi_4\,,\Phi_2))
- L_2^{(1)}(\Phi_4\,,\frac{b_0^+b_0^-}{L^+_0}\mathcal{P}c_0^-L_2^{(1)}(\Phi_2\,,\Phi_3))
\Big)\Bigg)
\nonumber\\
=&\ \omega_s\Bigg(\Phi_1\,,\Big(
L_3^{(2)}(\Phi_2\,,\Phi_3\,,\Phi_4)
- L_2^{(1)}(\Phi_2\,,\frac{b_0^+}{L^+_0}L_2^{(1)}(\Phi_3\,,\Phi_4))
\nonumber\\
&\hspace{8mm}
- L_2^{(1)}(\Phi_3\,,\frac{b_0^+}{L^+_0}L_2^{(1)}(\Phi_4\,,\Phi_2))
- L_2^{(1)}(\Phi_4\,,\frac{b_0^+}{L^+_0}L_2^{(1)}(\Phi_2\,,\Phi_3))
\Big)\Bigg)\,.
\end{align}
The second equality follows from the fact that the string field
$L_2^{(1)}(\Phi_1,\Phi_2)$ satisfies the closed string constraints
in Eq.\,(\ref{restrict closed}).
Using the coderivations this can also be written as
\begin{equation}
 \langle\mathcal{A}_4|\ =\ \langle\omega_s|\id\otimes
\pi_1^0\Big(\bd{L}_3^{(2)}|^0_0-\bd{L}_2^{(1)}|^0_0\,
\frac{b_0^+}{L^+_0}\,\bd{L}_2^{(1)}|^0_0\Big)\,.
\label{4NS 1}
\end{equation}
We show that Eq.\,(\ref{4NS 1}) actually agrees with
the first-quantized expression in Eq.\,(\ref{4NS 1st coderivation}).

First, we note that $\bd{L}|^0=\bd{B}|^0$ since $\hat{\bd{F}}=\id$ in the pure
NS sector. Substituting the relations
\begin{equation}
\bd{B}_2^{(1)}|^0_0\ =\ [\,\bd{Q}\,,\bd{\lambda}_2^{(1)}|^0_0\,]\,,\qquad
 2\bd{B}_3^{(2)}|^0_0\ =\ [\,\bd{Q}\,,\bd{\lambda}_3^{(2)}|^0_0\,] 
+ [\,\bd{B}^{(1)}_2|^0_0\,,\bd{\lambda}_2^{(1)}|^0_0\,]\,,
\end{equation}
following from Eq.\,(\ref{B general}) into Eq.\,(\ref{4NS 1}), we have
\begin{align}
 \langle\mathcal{A}_4|\ =\ \frac{1}{2}\langle\omega_s|\id\otimes\pi_1^0\Big(
&[\,\bd{Q}\,,\bd{\lambda}_3^{(2)}|^0_0\,]
+[\,\bd{B}^{(1)}_2|^0_0\,,\bd{\lambda}_2^{(1)}|^0_0\,]
\nonumber\\
&\
-[\,\bd{Q}\,,\bd{\lambda}_2^{(1)}|^0_0\,]\,\frac{b_0^+}{L_0^+}\,\bd{B}_2^{(1)}|^0_0
-\bd{B}_2^{(1)}|^0_0\,\frac{b_0^+}{L_0^+}\,[\,\bd{Q}\,,\bd{\lambda}_2^{(1)}|^0_0\,]
\Big)\,.
\label{4NS 2}
\end{align}
Moving to the large Hilbert space using $\langle\omega_s|=\langle\omega_l|\xi\otimes\id$
following from Eq.\,(\ref{omega l to s}), 
one can pull $\bd{Q}$ out from Eq.\,(\ref{4NS 2}) and obtain
\begin{equation}
 \langle\mathcal{A}_4|\ 
=\ - \frac{1}{2}\langle\omega_l|X_0\otimes\pi_1^0\Big(
\bd{\lambda}_3^{(2)}|^0_0
-\bd{\lambda}_2^{(1)}|^0_0\,\frac{b_0^+}{L_0^+}\,\bd{B}_2^{(1)}|^0_0
-\bd{B}_2^{(1)}|^0_0\,\frac{b_0^+}{L_0^+}\,\bd{\lambda}_2^{(1)}|^0_0\Big)\,,
\label{4NS 3}
\end{equation}
except for the terms vanishing when they hit the states in $\mathcal{H}_Q^{NS}$\,.
Similarly, we can further rewrite Eq.\,(\ref{4NS 3}) as
\begin{align}
 \langle\mathcal{A}_4|\ =&\ \frac{1}{2}\langle\omega_l|\xi X_0\otimes\pi_1^0\Big(
[\,\bd{\eta}\,,\bd{\lambda}_3^{(2)}|^0_0\,]
-[\,\bd{\eta}\,,\bd{\lambda}_2^{(1)}|^0_0\,]\,\frac{b_0^+}{L_0^+}\,\bd{B}_2^{(1)}|^0_0
-\bd{B}_2^{(1)}|^0_0\,\frac{b_0^+}{L_0^+}\,[\,\bd{\eta}\,,\bd{\lambda}_2^{(1)}|^0_0\,]\Big)
\nonumber\\
=&\
\frac{1}{2}\langle\omega_l|\xi X_0\otimes\pi_1^0\Big(
\bd{B}_3^{(1)}|^0_0
-\bd{L}_2^{(0)}|^0_0\,\frac{b_0^+}{L_0^+}\,\bd{B}_2^{(1)}|^0_0
-\bd{B}_2^{(1)}|^0_0\,\frac{b_0^+}{L_0^+}\,\bd{L}_2^{(0)}|^0_0\Big)\,,
\end{align}
using the relations
\begin{equation}
[\bd{\eta},\bd{\lambda}_3^{(2)}|^0_0]=\bd{B}_3^{(1)}|^0_0\,,\qquad
[\bd{\eta},\bd{\lambda}_2^{(1)}|^0_0]=\bd{L}_2^{(0)}|^0_0\,, 
\end{equation}
following from Eq.\,(\ref{lambda general}),
except for the terms vanishing when they hit the states in the small Hilbert space.
We can repeat similar steps once more.
Again using
\begin{subequations}  
\begin{alignat}{4}
 &\bd{B}_2^{(1)}|^0_0\ =\ [\,\bd{Q}\,,\bd{\lambda}_2^{(1)}|^0_0\,]\,,\qquad&
 &\bd{B}_3^{(1)}|^0_0\ =\ [\,\bd{Q}\,,\bd{\lambda}_3^{(1)}|^0_0\,]
+ [\,\bd{L}_2^{(0)}|^0_0\,,\bd{\lambda}_2^{(1)}|^0_0\,]\,,\\
& [\,\bd{\eta}\,,\bd{\lambda}_3^{(1)}|^0_0\,]\ =\ 2\bd{L}_3^{(0)}|^0_0\,,\qquad&
& [\,\bd{\eta}\,,\bd{\lambda}_2^{(1)}|^0_0\,]\ =\ \bd{L}_2^{(0)}|^0_0\,,
\end{alignat}
\end{subequations}
we find that it agrees with the first-quantized amplitude in
Eq.\,(\ref{4NS 1st coderivation}):
\begin{align}
 \langle\mathcal{A}_4|\ =&\ -\frac{1}{2}\langle\omega_l|(X_0)^2\otimes\pi_1^0\Big(
\bd{\lambda}_3^{(1)}|^0_0 
- \bd{\lambda}_2^{(1)}|^0_0\,\frac{b_0^+}{L_0^+}\,\bd{L}_2^{(0)}|^0_0
- \bd{L}_2^{(0)}\,\frac{b_0^+}{L_0^+}\,\bd{\lambda}_2^{(1)}|^0_0\Big)
\nonumber\\
=&\
\langle\omega_s|(X_0)^2\otimes\pi_1^0\Big(
\bd{L}_3^{(0)}|^0_0 - \bd{L}_2^{(0)}|^0_0\,\frac{b_0^+}{L_0^+}\,\bd{L}_2^{(0)}|^0_0\Big)\,.
\end{align}

\subsection{Two-NS-two-R amplitude}

Similarly, let us show the agreement of the two-NS-two-R amplitude.
The first quantized amplitude is now expressed as the multi-linear map
\begin{equation}
 \langle\mathcal{A}_4|\,:\,\mathcal{H}_Q^R\otimes
\mathcal{H}_Q^R\wedge(\mathcal{H}_Q^{NS})^{\wedge2} \rightarrow\mathbb{C}\,,
\end{equation}
which can be rewritten using the coderivations as
\begin{equation}
 \langle\mathcal{A}_4^{1st}|\ =\ \langle\omega_s|X_0\otimes\pi_1^1\Big(
\bd{L}_3^{(0)}|^2_0
- \bd{L}_2^{(0)}|^2_0\,\frac{b_0^+}{L_0^+}\,\bd{L}_2^{(0)}|^0_0
- \bd{L}_2^{(0)}|^2_0\,\frac{b_0^+}{L_0^+}\,\bd{L}_2^{(0)}|^2_0\Big)\,.
\end{equation}
In the heterotic string field theory, on the other hand, the two-NS-two-R amplitude
can be calculated as
\begin{align}
 \langle\mathcal{A}_4|\ =&\ 
\langle\omega_s|\id\otimes\pi_1^1\Big(\bd{b}_3|^2_0
- \bd{b}_2|^2_0\,\frac{b_0^+}{L_0^+}\,\bd{L}_2|^0_0
- \bd{b}_2|^2_0\,\frac{b_0^+X}{L_0^+}\,\bd{L}_2|^2_0\Big)
\nonumber\\
=&\
\langle\omega_s|\id\otimes\pi_1^1\Big(\bd{B}_3^{(1)}|^2_0
+ \bd{L}_2^{(0)}|^2_0\,\Xi\bd{L}_2^{(0)}|^2_0
- \bd{L}_2^{(0)}|^2_0\,\frac{b_0^+}{L_0^+}\,\bd{B}_2^{(1)}|^0_0
- \bd{L}_2^{(0)}|^2_0\,\frac{b_0^+X}{L_0^+}\,\bd{L}_2^{(0)}|^2_0\Big)\,,
\end{align}
where we used $\bd{B}_n^{(0)}=\bd{L}_n^{(0)}$ and
\begin{equation}
 \pi_1^1\bd{b}_2\ =\ \bd{L}_2^{(0)}|^2_0\,,\qquad
 \pi_1^1\bd{b}_3\ =\ \bd{L}_3^{(0)}|^4_2 + \bd{B}_3^{(1)}|^2_0
+ \bd{L}_2^{(0)}|^2_0\,\Xi\bd{L}_2^{(0)}|^2_0
\label{b and B}
\end{equation}
following from the definition $\pi_1\bd{b}=\pi_1\bd{B}\hat{\bd{F}}$\,. 
Again, $\Xi\bd{L}_2^{(0)}$ and $\frac{b_0^+X}{L_0^+}\bd{L}_2^{(0)}$
are coderivations derived from $\Xi L_2^{(0)}$ and 
$\frac{b_0^+X}{L_0^+}L_2^{(0)}$\,, respectively.
Using the relations
\begin{subequations}  
\begin{align}
& \bd{B}_2^{(1)}|^0_0\ =\ [\,\bd{Q}\,,\bd{\lambda}_2^{(1)}|^0_0\,]\,,\qquad
\bd{B}_3^{(1)}|^2_0\ =\ [\,\bd{Q}\,,\bd{\lambda}_3^{(1)}|^2_0\,]
+ [\,\bd{L}_2^{(0)}|^2_0\,,\bd{\lambda}_2^{(1)}|^0_0]|^2_0\,,\\
& [\,\bd{\eta}\,,\bd{\lambda}_2^{(1)}|^0_0\,]\ =\ \bd{L}_2^{(0)}|^0_0\,,\qquad
 [\,\bd{\eta}\,,\bd{\lambda}_3^{(1)}|^2_0\,]\ =\ \bd{L}_3^{(0)}|^2_0\,,
\end{align}
\end{subequations}
and $X=[\bd{Q},\Xi]$\,,
we find the amplitude obtained from the heterotic string field theory
agrees with the first-quantized amplitude:
\begin{align}
 \langle\mathcal{A}_4|\ =&\ - \langle\omega_l|X_0\otimes\pi_1^1\Big(
\bd{\lambda}_3^{(1)}|^2_0
-\bd{L}_2^{(0)}|^2_0\,\frac{b_0^+}{L_0^+}\,\bd{\lambda}_2^{(1)}|^0_0
-\bd{L}_2^{(0)}|^2_0\,\frac{b_0^+\Xi}{L_0^+}\,\bd{L}_2^{(0)}|^2_0\Big)
\nonumber\\
=&\ \langle\omega_s|X_0\otimes\pi_1^1\Big(
\bd{L}_3^{(0)}|^2_0
-\bd{L}_2^{(0)}|^2_0\,\frac{b_0^+}{L_0^+}\,\bd{L}_2^{(0)}|^0_0
-\bd{L}_2^{(0)}|^2_0\,\frac{b_0^+}{L_0^+}\,\bd{L}_2^{(0)}|^2_0\Big)
\Big)\ =\ \langle\mathcal{A}_4^{1st}|\,.
\end{align}

\subsection{Four-R amplitude}

Finally the four-R amplitude obtained from the heterotic string field theory 
expressed as a map
\begin{equation}
 \langle\mathcal{A}_4|\ :\ \mathcal{H}_Q^R\otimes(\mathcal{H}_Q^R)^{\wedge3}\ \rightarrow\ \mathbb{C}
\end{equation}
can be calculated as
\begin{align}
 \langle\mathcal{A}_4|\ =&\ 
\langle\omega_s|\id\otimes\pi_1^1\Big(\bd{b}_3|^4_2
- \bd{b}_2|^2_0\,\frac{b_0^+}{L_0^+}\,\bd{L}_2|^2_2\Big)
\nonumber\\
=&\
\langle\omega_s|\id\otimes\pi_1^1\Big(\bd{L}_3^{(0)}|^4_2
- \bd{L}_2^{(0)}|^2_0\,\frac{b_0^+}{L_0^+}\,\bd{L}_2^{(0)}|^2_2\Big)
\end{align}
by using Eq.\,(\ref{b and B}). 
This is nothing but the four-R amplitude in the first-quantized
formulation.

\sectiono{Gauge-invariant action in WZW-like formulation}
\label{action in WZW}

So far we have constructed a complete gauge-invariant action 
for the heterotic string field theory based on the cyclic
$L_\infty$ algebra in the small Hilbert space.
In this section we also construct a WZW-like action by using
a field redefinition, and show its gauge invariance independently.

\subsection{Complete action and gauge transformation}
First, we note that if we project $\bd{B}(s,t)$ and $\bd{\lambda}(s,t)$ onto
the pure NS sector,
%%%   2019/10/2
\begin{align}
 \bd{B}(s,t)|^0\ \equiv&\ \sum_{m,n=0}^\infty s^m t^n \bd{B}^{(n)}_{m+n+1}|^0\
=\ \sum_{m=0}^\infty s^m \bd{B}^{[m]}(t)|^0\,,\\
 \bd{\lambda}(s,t)|^0\ \equiv&\ \sum_{m,n=0}^\infty s^m t^n
 \bd{\lambda}^{(n+1)}_{m+n+2}|^0\
=\ \sum_{m=0}^\infty s^m \bd{\lambda}^{[m]}(t)|^0\,,
\end{align}
the differential equation in Eqs.\,(\ref{diff}) reduces to
\begin{subequations}\label{EKS diff} 
 \begin{align}
  \partial_t\bd{L}(s,t)|^{0}\ =&\ [\bd{L}(s,t)|^0\,,\,\bd{\lambda}(s,t)|^0]\,,
\label{EKS1}\\
 [\bd{\eta}\,,\,\bd{\lambda}(s,t)|^0]\ =&\ \partial_s\bd{L}(s,t)|^0\,,
\label{EKS2}
 \end{align}
\end{subequations}
where $\bd{L}(s,t)|^0\equiv\bd{Q}|^0+\bd{B}(s,t)|^0$\,,
and then, from Eqs.\,(\ref{IJst}), 
$\bd{L}(s,t)|^0$ %$\bd{B}(s,t)|^0$ 
satisfies
\begin{align}
 [\bd{L}(s,t)|^0\,,\,\bd{L}(s,t)|^0]\ =\ 0\,,\qquad
 [\bd{\eta}\,,\,\bd{L}(s,t)|^0]\ =\ 0\,.
\label{EKS L}
\end{align} 
The differential equations in Eqs.\,(\ref{EKS diff}) and $L_\infty$ relations 
in Eqs.\,(\ref{EKS L})
are nothing but those introduced in Ref.\cite{Erler:2014eba}.
Thus, by construction, the string and gauge products restricted in the pure NS
sector, 
$\bd{B}(s,t)|^0$ and $\bd{\lambda}(s,t)|^0$\,,
reduce to those in Ref.\cite{Erler:2014eba}.
This implies that the $L_\infty$ algebra restricting
Eq.\,(\ref{L hetero}) in the pure NS sector $\bd{Q}+\bd{B}^{[0]}|^0$  
can be written in the form 
of the similarity transformation, % of $\bd{Q}$\,,
\begin{equation}
%\hat{\bd{g}}\,\bd{L}|^{0}\,\hat{\bd{g}}^{-1}\
%=\ \bd{Q}\,,
%\bd{L}|^{0}\
\bd{Q}+\bd{B}^{[0]}|^0
=\ \hat{\bd{g}}^{-1}\,\bd{Q}\,\hat{\bd{g}}\,,
\label{L in NS}
\end{equation}
generated by the cohomomorphism \cite{Goto:2015pqv}
\begin{equation}
 \hat{\bd{g}}\ =\ 
\vec{\mathcal{P}}\exp\left(\int^1_0 dt\,\bd{\lambda}^{[0]}(t)|^0\right)\,.
\label{cohomo g}
\end{equation}
Here, $\vec{\mathcal{P}}$ %in (\ref{cohomo g}) 
denotes the path-ordered product from left to right.
Using this fact we find that the string products
$\bd{L}$ are transformed by (the inverse of) this similarity transformation as 
\begin{equation}
 \pi_1\tilde{\bd{L}}\ \equiv\ \pi_1\hat{\bd{g}}\,\bd{L}\,\hat{\bd{g}}^{-1}\
=\ \pi_1\bd{Q}+\pi_1^0\tilde{\bd{b}} + X \pi_1^1 \tilde{\bd{b}}\,,
\label{L tilde}
\end{equation}
where 
\begin{equation}
\tilde{\bd{b}}\ =\ 
  \hat{\bd{g}}\,(\bd{b}-\bd{B}^{[0]}|^0)\,\hat{\bd{g}}^{-1}\,.
\label{b tilde}
\end{equation}
By construction, $\bd{\lambda}^{[0]}(t)|^0$ is cyclic with respect to $\omega_l$\,.
This implies that the similarity transformation generated by
$\hat{\bd{g}}$ preserves the cyclicity, and thus $\tilde{\bd{b}}$ is also cyclic 
with respect to $\omega_l$.

Next, we rewrite the action in Eq.\,(\ref{small action}) in the WZW-like form
by extending the NS string field $\Phi_{NS}$ to $\Phi_{NS}(t)$ with $t\in[0,1]$ 
satisfying $\Phi_{NS}(1)=\Phi_{NS}$ and $\Phi_{NS}(0)=0$\,.
Using the cyclicity we find
\begin{align}
 S\ 
=&\
\frac{1}{2}\,\omega_s\big(\Phi_R\,,YQ\Phi_R\big)
+ \sum_{n=0}^\infty\frac{1}{(n+2)!}\,
\omega_s\big(\Phi_{NS}\,,L_{n+1}({\Phi_{NS}}^{n+1})\big)
\nonumber\\
&\
+ \sum_{n=1}^\infty\sum_{r=0}^\infty\,\frac{1}{(n+2r+2)(n-1)!(2r+2)!}\,
\omega_s\big(\Phi_{NS}\,,\,b_{n+2r+1}({\Phi_{NS}}^{n-1}\,,\,{\Phi_R}^{2r+2})\big)
\nonumber\\
&\
+ \sum_{n=0}^\infty\sum_{r=0}^\infty\,\frac{1}{(n+2r+2)n!(2r+1)!}\,
\omega_s\big(\Phi_{R}\,,\,b_{n+2r+1}({\Phi_{NS}}^{n}\,,\,{\Phi_R}^{2r+1})\big)
\nonumber\\
%%%%%%%%%%%
=&\ 
 \sum_{n=0}^\infty\frac{1}{(n+1)!}\,\int^1_0dt\,
\omega_s\big(\partial_t\Phi_{NS}(t)\,,\,L_{n+1}({\Phi_{NS}(t)}^{n+1})\big)
\nonumber\\
&\ 
+ \frac{1}{2}\,\omega_s\big(\Phi_R\,,YQ\Phi_R\big)
+\sum_{n=0}^\infty\sum_{r=0}^\infty
\frac{1}{n!(2r+2)!}\,\omega_s\big(\Phi_R\,,\,
b_{n+2r+1}({\Phi_{NS}}^n\,,\,{\Phi_R}^{2r+1})\big)
\nonumber
%%%%%%%%%%%%
%\nonumber\\
%&\
\end{align}
\begin{align}
=&\ \int^1_0 dt\, \omega_l\big(\xi\partial_t\Phi_{NS}(t)\,,
\pi_1^0\bd{L}(e^{\wedge\Phi_{NS}(t)})\big)
\nonumber\\
&\
+ \frac{1}{2}\,\omega_s(\Phi_R\,,YQ\Phi_R)
+ \sum_{r=0}^\infty\frac{1}{(2r+2)!}\,\omega_s\big(\Phi_R\,,\,
\pi_1^1\bd{b}(e^{\wedge\Phi_{NS}}\wedge\,{\Phi_R}^{\wedge 2r+1}\big)\,.
\label{WZW action 1}
\end{align}
Here, the first term can be mapped to the WZW-like action
as follows \cite{Goto:2015pqv}. 
Using the similarity transformation in Eq.\,(\ref{L tilde}) and the identity 
\begin{equation}
 \omega_l(\pi_1\hat{\bd{g}}\bd{l}_1(e^{\wedge\Phi})\,,
\pi_1\hat{\bd{g}}\bd{l}_2(e^{\wedge\Phi}))\ 
=\
\omega_l(\pi_1\bd{l}_1(e^{\wedge\Phi})\,,
\pi_1\bd{l}_2(e^{\wedge\Phi}))\,,
\label{id cyclic}
\end{equation}
for odd coderivations $\bd{l}_1$ and $\bd{l}_2$\,,
which we prove in Appendix \ref{id cyclic proof}\,,
the first term can be written as
\begin{equation}
\int^1_0 dt\, \omega_l\big(\xi\partial_t\Phi_{NS}(t)\,,
\pi_1^0\bd{L}(e^{\wedge\Phi_{NS}(t)})
\big)
=\
 \int^1_0 dt\, 
\omega_l\big(\pi_1\hat{\bd{g}}\bd{\xi_t}(e^{\wedge\Phi_{NS}(t)})\,,\,
\pi_1^0\bd{Q}\hat{\bd{g}}(e^{\wedge\Phi_{NS}(t)})\big)\,,
\label{WZW NS1}
\end{equation}
where $\bd{\xi_t}$ is the one coderivation derived from
$\xi\partial_t$\,.
By this transformation, the constraint 
$\eta\Phi=0$ restricting $\Phi$ in the
small Hilbert space is mapped to the constraint
\begin{equation}
0\ =\ \pi_1\hat{\bd{g}}\bd{\eta}(e^{\wedge(\Phi_{NS}+\Phi_R)})\ 
=\ \pi_1^0
\bd{L}^\eta(e^{\wedge \pi_1^0\hat{\bd{g}}(e^{\wedge\Phi_{NS}})})
+ \pi_1^1\eta\Phi_R\,,
\label{WZW map}
\end{equation}
where $\bd{L}^\eta\equiv\hat{\bd{g}}\,\bd{\eta}\,\hat{\bd{g}}^{-1}$\,.
Since this $\bd{L}^\eta$ is nothing but 
the dual $L_\infty$ products in Ref.\cite{Goto:2016ckh}, 
the NS component of Eq.\,(\ref{WZW map}) is the Maurer-Cartan equation for the pure-gauge 
string field $G_\eta(V)$ in the WZW-like formulation:
\begin{equation}
 \bd{L}^\eta(e^{\wedge G_\eta(V)})\ =\ 0\,.
\end{equation}
This suggests that we can identify the string fields 
$(\Phi_{NS},\Phi_R)$ with the string fields $(V,\Psi)$ in the WZW-like
formulation through the relations 
\begin{equation}
\pi_1^0\hat{\bd{g}}(e^{\wedge\Phi_{NS}})\ 
%g(\Phi_{NS})\ 
=\ G_\eta(V)\,,
\qquad
\Phi_R\ =\ \Psi\,.
\label{map NS}
\end{equation}
Then the associated fields $B_d(V(t))$ ($d=t,\delta$) are written as
\begin{equation}
B_d(V(t))\ =\ \pi_1^0\hat{\bd{g}}\bd{\xi_d}(e^{\wedge\Phi_{NS}(t)})
\label{associated field}
\end{equation}
under this identification,
where $\bd{\xi_\delta}$ is the one coderivation $\bd{\xi_\delta}$ 
derived from $\xi\delta$ \cite{Goto:2015pqv}.
The identities characterizing the associated field,
%%%   2019/10/22
\begin{subequations} \label{var id} 
\begin{align}
& dG_\eta(V(t))\ =\ \pi_1^0\bd{L}^\eta(e^{\wedge G_\eta(t)}\wedge B_d(V(t)))\,,
\label{key 1}\\
& D_\eta(t)\Big(
\partial_tB_\delta(V(t)) - \delta B_t(V(t)) +
\pi_1^0\bd{L}^\eta(e^{\wedge G_\eta(t)}\wedge\, B_t(V(t))\wedge\,B_\delta(V(t)))\Big)\ =\ 
0\,,
%D_\eta(\cdots)\,,
\label{key 2}
\end{align}
\end{subequations}
follow from the identifications in Eqs.\,(\ref{map NS}) and (\ref{associated field})
and the relation (\ref{relation group-like 2}), where we introduced the nilpotent linear operator 
$D_\eta(t)$ as
\begin{equation}
 D_\eta(t) \varphi\ =\ \pi_1^0\bd{L}^\eta(e^{\wedge G_\eta(t)}\wedge \varphi)\,,
%\quad \textrm{for}\ \varphi\in\mathcal{H}_{large}^{NS}\,.
\end{equation}
for a general string field $\varphi\in\mathcal{H}_{large}^{NS}$\,.
Eventually the first term of the action in Eq.\,(\ref{WZW action 1}) is mapped to
(the dual form of) the WZW-like action in the pure NS sector:
\begin{equation}
\int^1_0 dt\, \omega_l\big(\xi\partial_t\Phi_{NS}(t)\,,
\pi_1^0\bd{L}(e^{\wedge\Phi_{NS}(t)})\big)
=\
 \int^1_0 dt\, 
\omega_l\big(B_t(V(t))\,,\,QG_\eta(V(t))\big)\,.
\label{WZW NS2}
\end{equation}
If we note that $\pi_1^1\bd{b}=\pi_1^1\tilde{\bd{b}}\hat{\bd{g}}$\,,
the whole action is finally 
mapped to the complete WZW-like action\footnote{
The last term can also be written as
$\sum_{r=0}^\infty\frac{1}{(2r+2)!}\,\omega_l\big(\Psi\,,\,
\pi_1^1\hat{\bd{F}}_g(e^{\wedge G_\eta(V)}\wedge\Psi^{\wedge2r+1})\big)$
with $\hat{\bd{F}}_g=\hat{\bd{g}}\hat{\bd{F}}\hat{\bd{g}}^{-1}$
by using the relation 
$\pi_1\hat{\bd{F}}_g\ =\ \pi_1(\idSH + \Xi\pi_1^1\tilde{\bd{b}})$\,.
This form can be seen
as a natural extension of the result in Ref.\cite{Goto:2016ckh,Kunitomo-HRI,Kunitomo:2019did}. 
}
\begin{align}
S\ =&\ 
\int^1_0 dt\,\omega_l\big(B_t(V(t))\,,\,QG_\eta(V(t))\big) 
\nonumber\\
&\
+ \frac{1}{2}\,\omega_s(\Psi\,,\,YQ\Psi)
+ \sum_{r=0}^\infty\frac{1}{(2r+2)!}\,\omega_s\big(\Psi\,,\,
\pi_1^1\tilde{\bd{b}}(e^{\wedge G_\eta(V)}\wedge\Psi^{\wedge2r+1 })\big)\,.
\label{WZW action}
\end{align}

Since the field redefinition in Eq.\,(\ref{map NS}) does not uniquely determine 
the NS string field $V$\,, we have an extra gauge invariance in the WZW-like
formulation, 
\begin{equation}
B_\delta(V)=D_\eta\Omega\,,\qquad
\delta\Psi\ =\ 0\,,
\label{gauge tf 1}
\end{equation}
which keeps $G_\eta(V)$ invariant.
In addition, the gauge transformation in the small Hilbert space formulation,
\begin{equation}
 \pi_1\delta(e^{\wedge(\Phi_{NS}+\Phi_R)})\ =\ 
\pi_1\bd{L}(e^{\wedge(\Phi_{NS}+\Phi_R)}\wedge\,(\Lambda_{NS}+\Lambda_R))\,,
\end{equation}
is mapped to the gauge transformation in the WZW-like formulation,
except for the terms which can be absorbed into the transformation in 
Eq.\,(\ref{gauge tf 1}), as
\begin{subequations} \label{WZW gauge tf}
\begin{align}
  B_\delta(V)\ =&\ 
\pi_1^0\tilde{\bd{L}}(e^{\wedge(G_\eta+\Psi)}\wedge(\Lambda-\xi\lambda))\
=\ Q\Lambda 
+ \pi_1^0\tilde{\bd{b}}(e^{\wedge(G_\eta+\Psi)}\wedge(\Lambda-\xi\lambda))\,,\\
%%%%%%
 \delta\Psi\ =&\ 
\eta\pi_1^1\tilde{\bd{L}}(e^{\wedge(G_\eta+\Psi)}\wedge(\Lambda-\xi\lambda))\
=\ Q\lambda 
+ X\eta\pi_1^1\tilde{\bd{b}}(e^{(G_\eta+\Psi)}\wedge(\Lambda-\xi\lambda))\,,
\end{align}
\end{subequations}
with the identification of gauge parameters
\begin{equation}
\Lambda\ =\ - \pi_1^0\hat{\bd{g}}(e^{\wedge\Phi_{NS}}\wedge\xi\Lambda_{NS})\,,
\qquad
\lambda\ =\ \Lambda_R\,.
\end{equation}

\subsection{Gauge invariance}
\label{invariance}

Although it should be guaranteed by construction, we
prove here that the WZW-like action in Eq.\,(\ref{WZW action}),
\begin{align}
S\ =&\ 
\int^1_0 dt\,\omega_l\big(B_t(V(t))\,,\,QG_\eta(V(t))\big) 
\nonumber\\
&\
+ \frac{1}{2}\,\omega_s(\Psi\,,\,YQ\Psi)
%+ \sum_{r=0}^\infty\frac{1}{(2r+2)!}\,\omega_l\big(\Psi\,,\,
%\pi_1^1\hat{\bd{F}}_g(e^{\wedge G_\eta(V)}\wedge\Psi^{\wedge2r+1})\big)\,.
+ \sum_{r=0}^\infty\frac{1}{(2r+2)!}\,\omega_s\big(\Psi\,,\,
\pi_1^1\tilde{\bd{b}}(e^{\wedge G_\eta(V)}\wedge\Psi^{\wedge 2r+1})\big)\,.
\nonumber
\end{align}
is invariant under the gauge transformation in Eq.\,(\ref{WZW gauge tf}),
\begin{align}
 B_\delta(V)\ =&\ 
\pi_1^0\tilde{\bd{L}}\big(e^{\wedge(G_\eta+\Psi)}\wedge(\Lambda-\xi\lambda)\big)\,,
\nonumber\\
%%%%%%
 \delta\Psi\ =&\ 
\eta\pi_1^1\tilde{\bd{L}}(e^{\wedge(G_\eta+\Psi)}\wedge(\Lambda-\xi\lambda))\,,
\nonumber
\end{align}
in the WZW-like formulation independently. 

Let us first consider an arbitrary variation of the action.
In particular, the variation of the last term becomes
\begin{align}
\sum_{r=0}^\infty\frac{1}{(2r+2)!}\,\delta\omega_l\big(\xi\Psi\,,\,&
\pi_1^1\tilde{\bd{b}}(e^{\wedge G_\eta(V)}\wedge\Psi^{\wedge 2r+1})\big)
\nonumber\\
=&\
\sum_{r=0}^\infty\frac{1}{(2r+2)!}\,\omega_l\big(\xi\delta\Psi\,,\,
\pi_1^1\tilde{\bd{b}}(e^{\wedge G_\eta(V)}\wedge\Psi^{\wedge 2r+1})\big)
\nonumber\\
&\
- \sum_{r=0}^\infty\frac{2r+1}{(2r+2)!}\,\omega_l\big(\delta\Psi\,,\,
\pi_1^1\tilde{\bd{b}}(e^{\wedge G_\eta(V)}\wedge\Psi^{\wedge 2r}\wedge\,\xi\Psi)\big)
\nonumber\\
&\
- \sum_{r=0}^\infty\frac{1}{(2r+2)!}\,\omega_l\big(\delta G_\eta(V)\,,\,
\pi_1^0\tilde{\bd{b}}(e^{\wedge G_\eta(V)}\,\wedge\Psi^{\wedge 2r+1}\wedge\,\xi\Psi)\big)\,,
\label{var last}
\end{align} 
from the cyclicity of $\tilde{\bd{b}}$\,. 
The second term can further be calculated as
\begin{align}
- \sum_{r=0}^\infty\frac{2r+1}{(2r+2)!}\,\omega_l\big(\delta\Psi\,,\, &
\pi_1^1\tilde{\bd{b}}(e^{\wedge G_\eta(V)}\wedge\Psi^{\wedge 2r}\wedge\,\xi\Psi)\big)\
\nonumber\\ 
=&\
- \sum_{r=0}^\infty\frac{2r+1}{(2r+2)!}\,\omega_l\big(\xi\delta\Psi\,,\,
\pi_1^1\bd{L}^\eta\tilde{\bd{b}}(e^{\wedge G_\eta(V)}\wedge\Psi^{\wedge 2r}\wedge\,\xi\Psi)\big)\
\nonumber\\ 
=&\
 \sum_{r=0}^\infty\frac{2r+1}{(2r+2)!}\,\omega_l\big(\xi\delta\Psi\,,\,
\pi_1^1\tilde{\bd{b}}(e^{\wedge G_\eta(V)}\wedge\Psi^{\wedge 2r+1})\big)\,,
\end{align}
where we used $\pi_1^1\bd{\eta}=\pi_1^1\bd{L}^\eta$ and
$[\bd{L}^\eta,\tilde{\bd{b}}]=0$\,.
Adding up the first and second terms, the result is
\begin{equation}
 \sum_{r=0}^\infty\frac{1}{(2r+1)!}\,\omega_l\big(\xi\delta\Psi\,,\,
\pi_1^1\tilde{\bd{b}}(e^{(G_\eta})\wedge\,\Psi^{\wedge 2r+1}\big)\
=\
\omega_l\big(\xi\delta\Psi\,,\,\pi_1^1\tilde{\bd{b}}(e^{(G_\eta(V)+\Psi)})\big)\,.
\end{equation}
From the relation Eq.\,(\ref{key 1}),
the third term of Eq.\,(\ref{var last}) becomes
\begin{align}
%- &\sum_{r=0}^\infty\frac{1}{(2r+2)!}\,\omega_l\big(\delta G_\eta(V)\,,\,
%\pi_1^0\tilde{\bd{b}}(e^{\wedge G_\eta(V)}\,\wedge\Psi^{\wedge 2r+1}\wedge\,\xi\Psi)\big)\
%\nonumber\\
%=&\
- \sum_{r=0}^\infty\frac{1}{(2r+2)!}\,
\omega_l\big(\pi_1^0\bd{L}^\eta(e^{\wedge G_\eta(V)}\wedge\,B_\delta(V))\,,\,&
\pi_1\tilde{\bd{b}}(e^{\wedge G_\eta(V)}\,\wedge\Psi^{\wedge 2r+1}\wedge\,\xi\Psi)\big)\
\nonumber\\
=&\
 \sum_{r=0}^\infty\frac{1}{(2r+2)!}\,
\omega_l\big(B_\delta(V)\,,\,
\pi_1^0\tilde{\bd{b}}(e^{\wedge G_\eta(V)}\,\wedge\Psi^{\wedge 2r+2})\big)\
\nonumber\\
=&\
\omega_l\big(B_\delta(V)\,,\,
\pi_1^0\tilde{\bd{b}}(e^{\wedge (G_\eta(V)+\Psi)}\big)\,.
\end{align}
by using the fact that $\bd{L}^\eta$ is cyclic with respect to $\omega_l$
and it becomes $\bd{\eta}$ outside of the pure NS sector.
%except for acting in the pure NS sector.
In total, an arbitrary variation of the action becomes
\begin{align}
\delta S\ =&\ 
\omega_l\Big(B_\delta(V)\,,\,\big(QG_\eta + \pi_1^0\tilde{\bd{b}}(e^{(G_\eta(V)+\Psi)})\big)\Big)
+\omega_s\Big(\delta\Psi\,,\,Y\big(Q\Psi + X\pi_1^1\tilde{\bd{b}}(e^{(G_\eta(V)+\Psi)})\big)\Big)
\nonumber\\
=&\
\omega_l\Big(B_\delta(V)\,,\,\pi_1^0\tilde{\bd{L}}(e^{(G_\eta(V)+\Psi)})\Big)
+\omega_s\Big(\delta\Psi\,,\,Y\pi_1^1\tilde{\bd{L}}(e^{(G_\eta(V)+\Psi)})\Big)\,.
\end{align}
The equations of motion are therefore given by
\begin{equation}
 \pi_1\tilde{\bd{L}}(e^{\wedge(G_\eta(V)+\Psi)})\ =\ 0\,,
\end{equation}
which agrees with the equation
obtained by the similarity transformations of the equation of motion 
in the homotopy algebraic formulation:
\begin{equation}
 \pi_1\bd{L}(e^{\wedge(\Phi_{NS}+\Phi_R)})\ =\ 0\,.
\end{equation}

Now we can show %the gauge invariance of the action as
that the gauge transformation of the action becomes
\begin{align}
 \delta S\ 
%=&\
% \omega_l\Big(\pi_1^0\big(\bd{Q}\Lambda+\tilde{\bd{b}}
%(e^{\wedge(G_\eta+\Psi)}\wedge (\Lambda-\xi\lambda))\big)\,,
%\pi_1^0\tilde{\bd{L}}(e^{\wedge(G_\eta+\Psi)})
%\Big)
%\nonumber\\
%&\
%+ \omega_l\Big(\pi_1^1\big(-\bd{Q}\xi\lambda+X\tilde{\bd{b}}
%(e^{\wedge(G_\eta+\Psi)}\wedge (\Lambda-\xi\lambda))\big)\,,
%Y\pi_1^1\tilde{\bd{L}}(e^{\wedge(G_\eta+\Psi)})
%\Big)
%\nonumber\\
=&\
- \omega_l\Big(\pi_1^0\Lambda\,,
\pi_1^0\bd{Q}\tilde{\bd{L}}(e^{\wedge(G_\eta+\Psi)})\Big)
+ \omega_l\Big(\pi_1^0\tilde{\bd{b}}\big((\Lambda-\xi\lambda)\wedge\,
e^{\wedge(G_\eta+\Psi)}\big)\,,
\pi_1^0\tilde{\bd{L}}(e^{\wedge(G_\eta+\Psi)})\Big)
\nonumber\\
&\
+ \omega_l\Big(\pi_1^1\xi\lambda\,,
Y\pi_1^1\bd{Q}\tilde{\bd{L}}(e^{\wedge(G_\eta+\Psi)})\Big)
+ \omega_l\Big(\pi_1^1\tilde{\bd{b}}\big((\Lambda-\xi\lambda)\wedge\,
e^{\wedge(G_\eta+\Psi)}\big)\,,
\pi_1^1\tilde{\bd{L}}(e^{\wedge(G_\eta+\Psi)})\Big)
\nonumber\\
=&\
- \omega_l\Big(\pi_1^0\Lambda\,,
\pi_1^0\bd{Q}\tilde{\bd{L}}(e^{\wedge(G_\eta+\Psi)})\Big)
+ \omega_l\Big(\pi_1\xi\lambda\,,
Y\pi_1^1\bd{Q}\tilde{\bd{L}}(e^{\wedge(G_\eta+\Psi)})\Big)
\nonumber\\
&\
- \omega_l\Big(\pi_1(\Lambda-\xi\lambda)\,,\,
\pi_1\tilde{\bd{b}}\tilde{\bd{L}}(e^{\wedge(G_\eta+\Psi)})\Big)
\nonumber\\
%=&\
%- \omega_l\Big(\pi_1^0\Lambda\,,
%\pi_1^0(\bd{Q}+\tilde{\bd{b}})\tilde{\bd{L}}(e^{\wedge(G_\eta+\Psi)})\Big)
%+ \omega_l\Big(\pi_1\xi\lambda\,,
%Y\pi_1^1(\bd{Q}+X\tilde{\bd{b}})\tilde{\bd{L}}(e^{\wedge(G_\eta+\Psi)})\Big)
%\nonumber\\
=&\
- \omega_l\Big(\pi_1^0\Lambda\,,
\pi_1^0\tilde{\bd{L}}\tilde{\bd{L}}(e^{\wedge(G_\eta+\Psi)})\Big)
+ \omega_l\Big(\pi_1^1\xi\lambda\,,
Y\pi_1^1\tilde{\bd{L}}\tilde{\bd{L}}(e^{\wedge(G_\eta+\Psi)})\Big)
\nonumber\\
=&\ 0\,.
\end{align}
Hence the gauge invariance of the WZW-like action in Eq.\,(\ref{WZW action}) 
%under the gauge transformation (\ref{WZW gauge tf}) 
is shown 
%without refering the small Hilbert space formulation.
in the WZW-like formulation independently.

\sectiono{Summary and discussion}
\label{summary}

In this paper we have constructed a complete heterotic string field theory
in both the homotopy algebraic formulation and the WZW-like formulation. 
The complete action and gauge transformation in the homotopy algebraic 
formulation are given by means of string products realizing 
a cyclic $L_\infty$ algebra.
We have found that for constructing such string products it is useful 
to consider an $L_\infty$ algebra combining two $L_\infty$ algebras,
which can easily be cyclic.
Although these two, the dynamical and constraint 
$L_\infty$ algebras, are neither cyclic nor closed 
in the small Hilbert space, 
we can transform them by a similarity transformation
to the desired cyclic $L_\infty$ algebra, and
the constraint $L_\infty$ algebra restricting it in the small Hilbert space.
A concrete expression of the string products has been given by recursively
solving differential equations for their generating functions. 
It has been confirmed that this heterotic string field theory
reproduces all the physical four-point amplitudes at the tree level.
The WZW-like action and gauge transformation have also been 
constructed from those in the homotopy algebraic formulation through 
a field redefinition.

The remaining missing pieces of the homotopy algebraic and the WZW-like formulations
are complete type II superstring field theories. 
The prescription proposed in this paper can straightforwardly
be extended to this case. We can construct string products realizing
a cyclic $L_\infty$ algebra and complete gauge-invariant actions
in a similar manner. It will be reported in a separate paper \cite{Kunitomo-Sugimoto}.

Finally, needless to say that the construction of 
a complete action is not the end of
the story but just the beginning. We hope that the string field theory
constructed in this paper provides a useful basic approach
for studying various interesting nonperturbative properties of heterotic
strings.

\bigskip
\noindent
{\bf \large Acknowledgments}

The work of H.K. was supported in part by JSPS Grant-in-Aid for Scientific 
Research (C) Grant Number JP18K03645.

\appendix

\sectiono{Coalgebraic representation of $L_\infty$ algebra}
\label{coalgebra}

In this appendix we summarize basic definitions and properties of 
the coalgebraic representation of the $L_\infty$ algebra. 

Since the multi-closed-string products are
graded symmetric under interchange of their arguments, 
it is useful to introduce the symmetrized tensor product by
\begin{equation}
 \Phi_1\wedge\Phi_2\wedge\cdots\wedge\Phi_n\ =\
\sum_\sigma(-1)^{\epsilon(\sigma)}\Phi_{\sigma(1)}\otimes
\Phi_{\sigma(2)}\otimes\cdots\otimes\Phi_{\sigma(n)}\,,
\qquad \Phi_i\in\mathcal{H}\,,
\end{equation}
where $\sigma$ and $\epsilon(\sigma)$ denote all the permutations 
of $\{1,\cdots,n\}$
and the sign factor coming from the exchange $\{\Phi_1,\cdots,\Phi_n\}$
to $\{\Phi_{\sigma(1)},\cdots,\Phi_{\sigma(n)}\}$\,, respectively.
The string product $L_n(\Phi_1,\cdots,\Phi_n)$
can be represented as a linear operator $L_n$ which
maps the symmetrized tensor product of $n$ copies of $\mathcal{H}$ 
into $\mathcal{H}$\,,
\begin{equation}
 L_n\,:\, \mathcal{H}^{\wedge n}\ \rightarrow\ \mathcal{H}\,,
\end{equation}
defined by
\begin{equation}
  L_n(\Phi_1\wedge\cdots\wedge \Phi_n)\ =\ L_n(\Phi_1,\cdots,\Phi_n)\,.
\end{equation}
In order to consider an infinite sequence of these multi-string products
acting on different numbers of string fields, it is further useful
to introduce the symmetrized tensor algebra generated by 
%symmetrized tensor products of 
$\mathcal{H}$ as
\begin{equation}
 \mathcal{SH}\ =\ \mathcal{H}^{\wedge 0} \oplus \mathcal{H} \oplus
\mathcal{H}^{\wedge 2}\oplus \mathcal{H}^{\wedge 3}\oplus\cdots
\end{equation}
Here, $\mathcal{H}^{\wedge 0}$ is a one-dimensional vector space
spanned by the identity $\idSH$ of the symmetrized tensor product
satisfying
\begin{equation}
 \idSH\wedge V\ =\ V
\end{equation}
for any element $V\in\mathcal{SH}$\,.
The coderivation $\bd{L}_n$ is defined as an operator acting on $\mathcal{SH}$ by
\begin{subequations}\label{coderivationL}
 \begin{align}
 \bd{L_n}\,\Phi_1\wedge\cdots\wedge \Phi_m\ =&\ 0\,, \qquad &\textrm{for}\ m<n\,,\\
 \bd{L_n}\,\Phi_1\wedge\cdots\wedge \Phi_m\ =&\ L_n(\Phi_1\wedge\cdots\wedge\Phi_m)\,, 
\qquad &\textrm{for}\ m=n\,,\\
 \bd{L_n}\,\Phi_1\wedge\cdots\wedge \Phi_m\ =&\ \left(
L_n\wedge\id_{m-n}\right)
\,\Phi_1\wedge\cdots\wedge \Phi_m\,, \qquad &\textrm{for}\ m<n\,,
\end{align}
\end{subequations}
with
\begin{equation}
 \id_n\ =\ \frac{1}{n!}\,\underbrace{\id\wedge\cdots\wedge\id}_n\ 
=\ \underbrace{\id\otimes\cdots\otimes\id}_n\,.
\end{equation}
Then we can consider an infinite sequence of multi-string products collectively
as a general coderivation by adding them as
\begin{equation}
 \bd{L}\ =\ \bd{L}_1+\bd{L}_2+\bd{L}_3+\cdots\ =\
\sum_{n=0}^\infty \bd{L}_{n+1}\,.
\end{equation}
The $L_\infty$ relations in Eq.\,(\ref{L infinity}) are
represented as nilpotency of the coderivation $\bd{L}$\,:
\begin{equation}
 [\bd{L}\,,\,\bd{L}]\ =\ 0\,,
\end{equation}
where the square bracket denotes the \textit{graded} commutator,
and the coderivation $\bd{L}$ is assumed to be degree odd.

The $L_\infty$ structure is preserved under the $L_\infty$ isomorphism
represented by invertible cohomomorphisms.
A cohomomorphism is characterized by a sequence of degree-even multi-string 
products\footnote{
The zero-string product $H_0$ is also allowed in general, 
but we only consider the case with $H_0=0$ in this paper for simplicity.} 
\begin{equation}
 H_1(\Phi)\,,\, H_2(\Phi_1,\Phi_2)\,,\,H_3(\Phi_1,\Phi_2,\Phi_3)\,,\,\cdots\,,
\end{equation}
and defined by a linear operator on $\mathcal{SH}$\,,
\begin{equation}
 \hat{\bd{H}}\ =\ \pi_0 + \sum_{l=1}^\infty\,\frac{1}{l!}
\sum_{k_1,\cdots,k_l=1}^\infty
(H_{k_1}\wedge\cdots\wedge H_{k_l})\pi_{k_1+\cdots+k_l}\,,
\label{cohomo def}
%\label{cohomoL}
\end{equation}
where $\pi_n$ is the projection operator onto $\mathcal{H}^{\wedge n}$\ :
\begin{equation}
 \pi_n\,:\, \mathcal{SH}\ \rightarrow\ \mathcal{H}^{\wedge n}\,.
\label{projection 1}
\end{equation}
A useful relation obtained from the definition in Eq.\,(\ref{cohomo def}) is
\begin{align}
 & \hat{\bd{H}}(\Phi_1\wedge\cdots\wedge\Phi_k\wedge\Phi_a)
\nonumber\\
&=\ \sum_{p=0}^k\sum_\sigma (-1)^{\epsilon(\sigma)}
\hat{\bd{H}}(\Phi_{\sigma(1)}\wedge\cdots\wedge\Phi_{\sigma(p)})\,\wedge
\pi_1\hat{\bd{H}}(\Phi_{\sigma(p+1)}\wedge\cdots\wedge\Phi_{\sigma(k)}\wedge\Phi_a)\,,
\label{formula1}
\end{align}
or equivalently,
\begin{align}
 & \hat{\bd{H}}(\Phi_a\wedge\Phi_1\wedge\cdots\wedge\Phi_k)
\nonumber\\
&=\ \sum_{p=0}^k\sum_\sigma (-1)^{\epsilon(\sigma)}
\pi_1\hat{\bd{H}}(\Phi_a\wedge\Phi_{\sigma(1)}\wedge\cdots\wedge\Phi_{\sigma(p)})\wedge
\hat{\bd{H}}(\Phi_{\sigma(p+1)}\wedge\cdots\wedge\Phi_{\sigma(k)})\,,
\label{formula1-2}
\end{align}
where $\sigma$ denotes all the decompositions of $\{1,\cdots,k\}$ to
$\{\sigma(1)\,,\cdots\,,\sigma(p)\}$ and $\{\sigma(p+1)\,,\cdots\,,\sigma(k)\}$\,.
The symbol $\epsilon(\sigma)$ denotes the sign factor coming from the exchanging of 
the string fields due to this decomposition.
For an invertible cohomomorphism we have
\begin{align}
& \hat{\bd{H}}^{-1}(\Phi_a\,\wedge\,\hat{\bd{H}}(\Phi_1\,\wedge\,\cdots\,\wedge\,\Phi_k))\
\nonumber\\
&\
=\ \sum_{p=0}^{k}\sum_\sigma (-1)^{\epsilon(\sigma)}
\pi_1\hat{\bd{H}}^{-1}(\Phi_a\,\wedge\,
\hat{\bd{H}}(\Phi_{\sigma(1)}\,
\wedge\,\cdots\,\wedge\,\Phi_{\sigma(p)}))\,\wedge\,\Phi_{\sigma(p+1)}\,
\wedge\,\cdots\,\wedge\,
\Phi_{\sigma(k)}\,.
\label{formula2}
\end{align}

A group-like element is a useful object in the tensor algebra $\mathcal{SH}$\,,
and is defined by
\begin{equation}
 e^{\wedge\Phi} \equiv \idSH + \Phi + \frac{1}{2!}\Phi\wedge \Phi + 
\frac{1}{3!}\Phi\wedge \Phi\wedge \Phi + \cdots\
=\ \sum_{n=0}^\infty \frac{1}{n!}\Phi^{\wedge n}
\label{groupL}
\end{equation}
for a given Grassmann-even string field $\Phi$\,.
A coderivation $\bd{l}$ and a cohomomorphism $\hat{\bd{H}}$ act 
on the group-like element as
\begin{equation}
 \bd{l}(e^{\wedge\Phi})\ =\ e^{\wedge\Phi}\wedge\pi_1\bd{l}(e^{\wedge\Phi})\,,
\qquad
 \hat{\bd{H}}(e^{\wedge\Phi})\ =\ e^{H[\Phi]}\,,
\label{on group-like}
\end{equation}
where $ H[\Phi]=\pi_1\hat{\bd{H}}(e^{\wedge\Phi})$\,.
The relation
\begin{align}
 \hat{\bd{H}}(e^{\wedge\Phi}\wedge \Phi_a)\ =&\ \hat{\bd{H}}(e^{\wedge\Phi})\wedge
\pi_1\hat{\bd{H}}(e^{\wedge\Phi}\wedge\Phi_a)
\label{relation group-like}\\
 \hat{\boldsymbol{H}}(e^{\wedge\Phi}\wedge\Phi_a\wedge\Phi_b)\ =&\
\hat{\boldsymbol{H}}(e^{\wedge\Phi})\wedge\pi_1\hat{\boldsymbol{H}}(e^{\wedge\Phi}\wedge\Phi_a\wedge\Phi_b)
\nonumber\\
&
+\hat{\boldsymbol{H}}(e^{\wedge\Phi})\wedge\pi_1\hat{\boldsymbol{H}}(e^{\wedge\Phi}\wedge\Phi_a)\wedge
\pi_1\hat{\boldsymbol{H}}(e^{\wedge\Phi}\wedge\Phi_b)\,,
\label{relation group-like 2}
\end{align}
those follow from the definition are useful.

\sectiono{Counting Ramond states}
\label{counting R}

Let us summarize the ways to count the Ramond states in this appendix.

We introduce a projection operator $\pi^r$ in the symmetrized tensor algebra,
which projects onto the states containing
$r$ Ramond states:
\begin{equation}
 \pi^r\,:\, \mathcal{SH}\ \rightarrow\ 
\{\Phi_{R\, 1}\wedge\cdots\wedge\Phi_{R\,r}\wedge\Phi_{NS\,1}
\wedge\Phi_{NS\,2}\wedge\cdots\}\,.
\end{equation}
By multiplying this by the projection operator in Eq.\,(\ref{projection 1}), 
we can define a projection operator $\pi_n^r$ selecting states
containing $r$ Ramond states in $\mathcal{H}^{\wedge n}$\,:
\begin{equation}
 \pi_n^r\ =\ \pi_n\pi^r\,:\, \mathcal{SH}\ \rightarrow\ 
\{\Phi_{R\, 1}\wedge\cdots\wedge\Phi_{R\, r}\wedge
\Phi_{NS\,1}\wedge\cdots\wedge\Phi_{NS\, n-r}\}\,.
\end{equation}
Note that a coderivation corresponding to the $n$-string product $\bd{l}_n$
satisfies
\begin{equation}
 \pi_{m+1}\bd{l}_n\ =\ \bd{l}_n\pi_{m+n}\,.
\label{relation 1}
\end{equation} 
We can decompose the coderivation $\bd{l}_n$
by the Ramond number defined by the number of Ramond inputs 
minus the number of Ramond outputs,
which is denoted as the subscript after a vertical line,
like $\bd{l}_n|_{r}$\,: %\,. The coderivation $\bd{l}_n|_{r}$ satisfies
\begin{equation}
 \pi^{s}\bd{l}_n|_{r}\ =\ \bd{l}_n|_{r}\pi^{r+s}\,.
\label{relation 2}
\end{equation}
If we note that the NS and Ramond states represent the space-time bosons and
fermions, respectively, the Ramond number must be even. 
Therefore the range of the Ramond number is
\begin{equation}
 0 \le r \le 2 \left[\frac{n}{2}\right]\,,
\end{equation}
where $[x]$ is Gauss' symbol representing the greatest
integer that is less than or equal to $x$\,.
Combining Eqs.\,(\ref{relation 1}) and (\ref{relation 2}),
it is found that a coderivation $\bd{l}_n|_{r}$ satisfies
\begin{equation}
 \pi^s_{m+1}\bd{l}_n|_{r}\ =\ \bd{l}_n|_{r}\pi_{m+n}^{r+s}\,.
\label{relation 3}
\end{equation}
It is important to note that the Ramond number is additive when
taking the (graded) commutator:
\begin{equation}
 [\bd{l}_n|_r\,, \bd{l}'_m|_s]|_{r+s}\ =\ [\bd{l}_n|_r\,, \bd{l}'_m|_s]\,.
\end{equation}

In order to consider the cyclicity, the cyclic Ramond number
is more useful than the Ramond number because it is preserved 
under the cyclic permutation.
The cyclic Ramond number is defined by the sum of the numbers of 
Ramond inputs and Ramond outputs, which must also be even.
The cyclic Ramond number is denoted by a superscript after a vertical line,
and a coderivation $\bd{l}_n|^{r}$ satisfies
\begin{equation}
 \pi^s_{m+1}\bd{l}_n|^{r}\ =\ \bd{l}_n|^{r}\pi_{m+n}^{r-s}\,.
\end{equation}
The range of the cyclic Ramond number is
\begin{equation}
0 \le r \le 2 \left[\frac{n+1}{2}\right]\,.
\end{equation}
A coderivation with a definite Ramond number or with 
a definite cyclic Ramond number is decomposed as
\begin{align}
l_n|_{2r}\ =&\ \pi_1^0\,\bd{l}_n|_{2r}^{2r} + \pi_1^1\,\bd{l}_n|_{2r}^{2r+2}\,,
\label{R decomp}\\
l_n|^{2r}\ =&\ \pi_1^0\,\bd{l}_n|_{2r}^{2r} + \pi_1^1\,\bd{l}_n|_{2r-2}^{2r}\,.
\label{cyclic R decomp}
\end{align}
The cyclic Ramond number, however, is not additive when
taking a commutator: the commutator of coderivations with definite
cyclic Ramond numbers does not have a definite cyclic number.
From the decomposition in Eq.\,(\ref{cyclic R decomp}) we have
\begin{equation}
 [\,\bd{l}_n|^{2r}\,,\bd{l}'_m|^{2s}\,]\ =\ 
 [\,\bd{l}_n|^{2r}\,,\bd{l}'_m|^{2s}\,]\,|^{2r+2s} 
+ [\,\bd{l}_n|^{2r}\,,\bd{l}'_m|^{2s}\,]\,|^{2r+2s-2}\,.
\label{cyclic Ramond decomposition}
\end{equation}
So, it is useful to introduce operations picking up a part with definite cyclic 
Ramond numbers as follows.
Suppose that there are two coderivations $\bd{l}=\sum_r\bd{l}|^{2r}$ and
$\bd{l}'=\sum_s\bd{l}'|^{2s}$\,. We define two operations 
%$[\,,\,]^1$ and $[\,,\,]^2$ as
\begin{subequations}\label{two operations} 
\begin{align}
  [\,\bd{l}\,, \bd{l}'\,]^1\ =&\ 
\sum_{r,s}\,[\,\bd{l}\,|^{2r}\,, \bd{l}'\,|^{2s}\,]\,|^{2r+2s}\,,\\
 [\,\bd{l}\,, \bd{l}'\,]^2\ =&\ 
\sum_{r,s}\,[\,\bd{l}\,|^{2r}\,, \bd{l}'\,|^{2s}\,]\,|^{2r+2s-2}\,,
\end{align}
\end{subequations}
which decompose a commutator as
\begin{equation}
 [\,\bd{l}\,, \bd{l}'\,]\ =\ [\,\bd{l}\,, \bd{l}'\,]^1 + [\,\bd{l}\,, \bd{l}'\,]^2\,.
\label{commutator decomposition}
\end{equation}
We can show from Eq.\,(\ref{cyclic Ramond decomposition}) that
\begin{align}
[[\,\bd{l}\,|^{2r}\,, \bd{l}'\,|^{2s}\,]\,|^{2r+2s}\,, 
\bd{l}''\,|^{2t}\,]\,|^{2r+2s+2t}\ =&\
[[\,\bd{l}\,|^{2r}\,, \bd{l}'\,|^{2s}\,]\,, 
\bd{l}''\,|^{2t}\,]\,|^{2r+2s+2t}\,,\\
[[\,\bd{l}\,|^{2r}\,, \bd{l}'\,|^{2s}\,]\,|^{2r+2s-2}\,, 
\bd{l}''\,|^{2t}\,]\,|^{2r+2s+2t-4}\ =&\
[[\,\bd{l}\,|^{2r}\,, \bd{l}'\,|^{2s}\,]\,, \bd{l}''\,|^{2t}\,]\,|^{2r+2s+2t-4}\,.
\end{align}
These imply that the operations $[\ , \ ]^1$ and $[\ ,\ ]^2$
satisfy the Jacobi identities
\begin{align}
& [[\,\bd{l}\,, \bd{l}'\,]^1\,, \bd{l}''\,]^1 
+ (-1)^{l(l'+l'')}[[\,\bd{l}'\,, \bd{l}''\,]^1\,, \bd{l}\,]^1 
+ (-1)^{l'(l''+l)}[[\,\bd{l}''\,, \bd{l}\,]^1\,, \bd{l}'\,]^1\ =\ 0\,,\\
& [[\,\bd{l}\,, \bd{l}'\,]^2\,, \bd{l}''\,]^2 
+ (-1)^{l(l'+l'')}[[\,\bd{l}'\,, \bd{l}''\,]^2\,, \bd{l}\,]^2
+ (-1)^{l'(l''+l)}[[\,\bd{l}''\,, \bd{l}\,]^2\,, \bd{l}'\,]^2\ =\ 0\,,
\end{align}
where $\bd{l}''=\sum_t\bd{l}''\,|^{2t}$\,.
Due to the Jacobi identity of the commutator
and the decomposition in Eq.\,(\ref{commutator decomposition}), the identity
\begin{align}
 & [[\,\bd{l}\,, \bd{l}'\,]^1\,, \bd{l}''\,]^2 +  
[[\,\bd{l}\,, \bd{l}'\,]^2\,, \bd{l}''\,]^1 
\nonumber\\
&\hspace{10mm}
+ (-1)^{l(l'+l'')}[[\,\bd{l}'\,, \bd{l}''\,]^1\,, \bd{l}\,]^2 
+ (-1)^{l(l'+l'')}[[\,\bd{l}'\,, \bd{l}''\,]^2\,, \bd{l}\,]^1
\nonumber\\
&\hspace{20mm}
+ (-1)^{l''(l+l')}[[\,\bd{l}''\,, \bd{l}\,]^1\,, \bd{l}'\,]^2\ 
+ (-1)^{l''(l+l')}[[\,\bd{l}''\,, \bd{l}\,]^2\,, \bd{l}'\,]^1\ =\ 0\,,
\end{align}
also holds.
Note that in the pure NS sector, 
\begin{subequations} \label{operation in NS}
 \begin{align}
 [\,\bd{l}|^0\,, \bd{l}'|^0\,]^1\ =&\  [\,\bd{l}|^0\,, \bd{l}'|^0\,]|^0\
=\  [\,\bd{l}|^0\,, \bd{l}'|^0\,]\,,\\
 [\bd{l}|^0\,, \bd{l}'|^0]^2\ =&\ 0\,.
\end{align}
\end{subequations}

\sectiono{A proof of cyclicity}
\label{cyclicity proof}

We first note that if $\bd{b}$ is cyclic with respect to $\omega_l$\,,
then $\bd{b}$ is cyclic with respect to $\omega_s$\,:
\begin{align}
  \omega_s(\Phi_1\,,b_n(\Phi_2\,,\cdots\,,\Phi_{n+1}))\ =&\
 \omega_l(\xi\Phi_1\,,b_n(\Phi_2\,,\cdots\,,\Phi_{n+1}))
\nonumber\\
=&\ (-1)^{|\Phi_1|}\omega_l(b_n(\xi\Phi_1\,,\Phi_2\,,\cdots\,,\Phi_n)\,,\Phi_{n+1})
\nonumber\\
=&\ -(-1)^{|\Phi_1|}\omega_l(\xi b_n(\Phi_1\,,\Phi_2\,,\cdots\,,\Phi_n)\,,\Phi_{n+1})
\nonumber\\
=&\ -(-1)^{|\Phi_1|}\omega_s(b_n(\Phi_1\,,\Phi_2\,,\cdots\,,\Phi_n)\,,\Phi_{n+1})\,.
\end{align}
Here, the second equality comes from the assumption, and 
we used $[\eta,b_n]=0$ after inserting $\{\eta,\xi\}=1$ in front of $b_n$
in the third equality. Therefore it is sufficient to prove that
if $\bd{B}$ is cyclic with respect to $\omega_l$, then 
$\pi_1\bd{b}=\pi_1\bd{B}\hat{\bd{F}}$ 
is also cyclic with respect to $\omega_l$\,.

We use mathematical induction with respect to the number of inputs. 
Since $b_2=B_2$\,, it is cyclic with respect to $\omega_l$ from the assumption.
Next assume that $b_n$ for $2\le n\le k$ is cyclic with respect to $\omega_l$\,.
Using the relation in Eq.\,(\ref{formula1}) we find that
\begin{align}
& \omega_l\Big(\varphi_a\,,
 b_{k+1}(\varphi_1\,\wedge\,\cdots\wedge\,\varphi_k\,\wedge\,\varphi_b)\Big)
=\ \omega_l\Big(\varphi_a\,,
 \pi_1\bd{B}\hat{\bd{F}}(\varphi_1\,\wedge\,\cdots\wedge\,\varphi_k\,\wedge\,\varphi_b)\Big)
\nonumber\\
&=\ \sum_{p=1}^{k}\sum_\sigma(-1)^{\epsilon(\sigma)}
\omega_l\Big(\varphi_a\,,\pi_1\bd{B}
\big(\hat{\bd{F}}(\varphi_{\sigma(1)}\,
\wedge\,\cdots\,\wedge\,\varphi_{\sigma(p)})\,\wedge\,\pi_1\hat{\bd{F}}(\varphi_{\sigma(p+1)}\,
\wedge\,\cdots\,\wedge\,\varphi_{\sigma(k)}\,\wedge\,\varphi_b)\big)\Big)
\nonumber\\
&=\ - \sum_{p=1}^{k}\sum_\sigma 
(-1)^{|\varphi_a|+\epsilon(\sigma)}
\omega_l\bigg(
\pi_1\bd{B}\Big(\varphi_a\,\wedge\,
\hat{\bd{F}}\big(\varphi_{\sigma(1)}\,\wedge\,\cdots\,\wedge\,\varphi_{\sigma(p)}\big)\Big)\,,\,
\pi_1\hat{\bd{F}}\big(\varphi_{\sigma(p+1)}\,
\wedge\,\cdots\,\wedge\,\varphi_{\sigma(k)}\,\wedge\,\varphi_b\big)
\bigg)
\nonumber\\
&=\ - (-1)^{|\varphi_a|}
 \omega_l\bigg(\pi_1\bd{B}\Big(\varphi_a\,\wedge\,\hat{\bd{F}}\big(\varphi_1\,\wedge\,
\cdots\,\wedge\,\varphi_k\big)\Big)\,,\,\varphi_b\bigg)
\nonumber\\
&\
- \sum_{p=1}^{k}\sum_\sigma (-1)^{|\varphi_a|+\epsilon(\sigma)}
\omega_l\bigg(
\pi_1^1\bd{B}\Big(\varphi_a\,\wedge\,
\hat{\bd{F}}\big(\varphi_{\sigma(1)}\,\wedge\,\cdots\,\wedge\,\varphi_{\sigma(p)}\big)\Big)\,,
\Xi\pi_1^1\bd{b}\,\big(\varphi_{\sigma(p+1)}\,
\wedge\,\cdots\,\wedge\,\varphi_{\sigma(k)}\,\wedge\,\varphi_b\big)\bigg)\,,
\label{proof1}
\end{align}
where in the last equality we used Eq.\,(\ref{defF}).

Using the cyclicity of $b_n$ for $n\le k$\,, the second term further becomes
\begin{align}
&\ - \sum_{p=1}^{k}\sum_\sigma (-1)^{|\varphi_a|+\epsilon(\sigma)}
\omega_l\bigg(
\pi_1^1\bd{B}\Big(\varphi_a\,\wedge\,
\hat{\bd{F}}\big(\varphi_{\sigma(1)}\,\wedge\,\cdots\,\wedge\,\varphi_{\sigma(p)}\big)\Big)\,,\,
\Xi\pi_1^1\bd{b}\,\big(\varphi_{\sigma(p+1)}\,
\wedge\,\cdots\,\wedge\,\varphi_{\sigma(k)}\,\wedge\,\varphi_b\big)
\bigg)
\nonumber\\
&=\ - \sum_{p=1}^{k}\sum_\sigma (-1)^{|\varphi_a|+\epsilon(\sigma)}
\omega_l\bigg(
\pi_1\bd{b}\,\Big(\Xi\pi_1^1\bd{B}\,\big(\varphi_a\,\wedge\,
\hat{\bd{F}}(\varphi_{\sigma(1)}\,\wedge\,\cdots\,\wedge\,\varphi_{\sigma(p)})\big)\,\wedge\,
\varphi_{\sigma(p+1)}\,\wedge\,\cdots\,\wedge\,\varphi_{\sigma(k)}\Big)\,,\,\varphi_b\bigg)
\nonumber\\
&=\ \sum_{p=1}^{k}\sum_\sigma (-1)^{|\varphi_a|+\epsilon(\sigma)}
\omega_l\bigg(
\pi_1\bd{B}\hat{\bd{F}}\,\Big(
\pi_1\hat{\bd{F}}^{-1}\,\big(\varphi_a\,\wedge\,
\hat{\bd{F}}\big(\varphi_{\sigma(1)}\,\wedge\,\cdots\,\wedge\,\varphi_{\sigma(p)})\big)\,\wedge\,
\varphi_{\sigma(p+1)}\,\wedge\,\cdots\,\wedge\,\varphi_{\sigma(k)}\Big)\,,\,
\varphi_b\bigg)\,.
\label{proof2}
\end{align}
Here, for $0<p<k$ the relation
\begin{align}
0\ =&\ \pi_1\hat{\bd{F}}^{-1}\hat{\bd{F}}\big(\varphi_a\,\wedge\,\varphi_{\sigma(1)}\,\wedge\,
\cdots\,\wedge\,\varphi_{\sigma(p)}\big)
\nonumber\\
=&\ \pi_1\hat{\bd{F}}^{-1}\Big(\varphi_a\,\wedge\,
\hat{\bd{F}}\big(\varphi_{\sigma(1)}\,\wedge\,\cdots\,\wedge\,\varphi_{\sigma(p)}\big)\Big)
\nonumber\\
&\
+ \sum_{q=0}^{p}\sum_\tau (-1)^{\epsilon(\tau)}
\pi_1\hat{\bd{F}}^{-1}\Big(\pi_1\hat{\bd{F}}(\varphi_a\,\wedge\,\varphi_{\tau(\sigma(1))}\,
\wedge\,\cdots\,\wedge\,\varphi_{\tau(\sigma(q))})
\nonumber\\
&\hspace{60mm}
\wedge\,\hat{\bd{F}}(\varphi_{\tau(\sigma(q+1))}\,\wedge\,\cdots\,\wedge\,
\varphi_{\tau(\sigma(p))})\Big)
\label{proof3}
\end{align}
holds. Then we have
\begin{align}
& \sum_{p=1}^{k}\sum_\sigma\ (-1)^{|\varphi_a|+\epsilon(\sigma)}
\pi_1\hat{\bd{F}}^{-1}\Big(\varphi_a\,\wedge\,
\hat{\bd{F}}\big(\varphi_{\sigma(1)}\,\wedge\,
\cdots\,\wedge\,\varphi_{\sigma(p)}\big)\Big)\,\wedge\,\varphi_{\sigma(p+1)}\,\wedge\,
\cdots\,\wedge\,\varphi_{\sigma(k)}\
\nonumber\\
&=\ - \sum_{p=1}^{k}\sum_\sigma\sum_{q=0}^{p}\sum_\tau 
(-1)^{|\varphi_a|+\epsilon(\sigma)+\epsilon(\tau)} 
\nonumber\\
&\hspace{20mm} \times
\pi_1\hat{\bd{F}}^{-1}\bigg(\pi_1\hat{\bd{F}}\Big(\varphi_a\,\wedge\,
\varphi_{\tau(\sigma(1))}\,\wedge\,
\cdots\,\wedge\,\varphi_{\tau(\sigma(q))}\Big)\,
\nonumber\\
&\hspace{50mm}
\wedge\,\hat{\bd{F}}\Big(\varphi_{\tau(\sigma(q+1)))}\,
\wedge\,\cdots\,\wedge\,\varphi_{\tau(\sigma(p))}\Big)\bigg)\,\wedge\,
\varphi_{\sigma(p+1)}\,\wedge\,\cdots\,
\wedge\,\varphi_{\sigma(k)}
\nonumber\\
&=\ -\sum_{p=1}^k\sum_{\sigma} (-1)^{|\varphi_a|+\epsilon(\sigma)}
\hat{\bd{F}}^{-1}\Big(\pi_1\hat{\bd{F}}\big(\varphi_a\,\wedge\,\varphi_{\sigma(1)}\,
\wedge\,\cdots\,\wedge\,\varphi_{\sigma(p)}\big)\,\wedge\,
\hat{\bd{F}}\big(\varphi_{\sigma(p+1)}\,\wedge\,\cdots\,\wedge\,\varphi_{\sigma(k)}
\big)\Big)\,,
\end{align}
where we used Eq.\,(\ref{formula2}) in the last equality.
Substituting this into the expression in Eq.\,(\ref{proof2}), we can show
that $b_{k+1}$ is cyclic with respect to $\omega_l$\,:
\begin{align}
& \omega_l\Big(\varphi_a\,,
 b_{k+1}(\varphi_1\,,\cdots\,,\varphi_k\,,\varphi_b)\Big) 
\nonumber\\
&=\
- (-1)^{|\varphi_a|}
 \omega_l\bigg(\pi_1\bd{B}\Big(\varphi_a\,\wedge\,\hat{\bd{F}}\big(\varphi_1\,\wedge\,
\cdots\,\wedge\,\varphi_k\big)\Big)\,,\,\varphi_b\bigg)
\nonumber\\
&\
- \sum_{p=1}^{k-1}\sum_\sigma (-1)^{|\varphi_a|+\epsilon(\sigma)}
\omega_l\bigg(
\pi_1^1\bd{B}\Big(\pi_1\hat{\bd{F}}\big(\varphi_a\,\wedge\,
\varphi_{\sigma(1)}\,\wedge\,\cdots\,\wedge\,\varphi_{\sigma(p)}\big)\,
\wedge\,\hat{\bd{F}}\big(\varphi_{\sigma(p+1)}\,
\wedge\,\cdots\,\wedge\,\varphi_{\sigma(k)}\big)\Big)\,,\,
\varphi_b\bigg)
\nonumber\\
&=\ 
- \sum_{p=0}^{k-1}\sum_\sigma (-1)^{|\varphi_a|+\epsilon(\sigma)}
\omega_l\bigg(
\pi_1^1\bd{B}\Big(\pi_1\hat{\bd{F}}\big(\varphi_a\,\wedge\,
\varphi_{\sigma(1)}\,\wedge\,\cdots\,\wedge\,\varphi_{\sigma(p)}\big)\,
\wedge\,\hat{\bd{F}}\big(\varphi_{\sigma(p+1)}\,
\wedge\,\cdots\,\wedge\,\varphi_{\sigma(k)}\big)\Big)\,,\,
\varphi_b\bigg)
\nonumber\\
&=\
-(-1)^{|\varphi_a|}
\omega_l\Big(\pi_1\bd{B}\hat{\bd{F}}(\xi\Phi_a\,\wedge\,\Phi_1\,\wedge\,
\cdots\,\wedge\,\Phi_k)\,,\,\Phi_b\Big)
\nonumber\\
&=\
-(-1)^{|\varphi_a|}
 \omega_l\Big(
b_{k+1}(\varphi_a\,, \varphi_1\,,\cdots\,,\varphi_k)\,,
\varphi_b\Big)\,.
\end{align}
Hence, $b_n$ is cyclic with respect to $\omega_l$ for arbitrary $n$\,.
That is, $\bd{b}$ is cyclic with respect to $\omega_l$\,.

\section{A proof of the identity in Eq.\,(\ref{id cyclic})}
\label{id cyclic proof}

%We prove the identity (\ref{id cyclic}).
Define
\begin{equation}
 \hat{\bd{g}}(t_0)\ =\vec{\mathcal{P}}\exp\left(
\int^{1}_{t_0}dt\bd{\lambda}(t)\right)\,,
\end{equation}
with $\bd{\lambda}(t)\equiv\bd{\lambda}^{[0]}(t)|^0$\,, 
which implies that
\begin{equation}
\partial_{t_0}\hat{\bd{g}}(t_0)\ 
=\ \bd{\lambda}(t_0)\hat{\bd{g}}(t_0)\,.  
\end{equation}
Using Eqs.\,(\ref{on group-like}) and (\ref{relation group-like}), 
we can show that
\begin{align}
&\partial_{t_0}\omega_l(\pi_1\hat{\bd{g}}(t_0)\bd{l}_1(e^{\wedge\Phi})\,,
\pi_1\hat{\bd{g}}(t_0)\bd{l}_2(e^{\wedge\Phi}))\ 
\nonumber\\
&=\
\omega_l(\pi_1\bd{\lambda}(t_0)\hat{\bd{g}}(t_0)
(e^{\wedge\Phi}\wedge\pi_1\bd{l}_1(e^{\wedge\Phi}))\,,
\pi_1\hat{\bd{g}}(t_0)(e^{\wedge\Phi}\wedge\pi_1\bd{l}_2(e^{\wedge\Phi})))
\nonumber\\
&\ \ \
+\omega_l(\pi_1\hat{\bd{g}}(t_0)
(e^{\wedge\Phi}\wedge\pi_1\bd{l}_1(e^{\wedge\Phi}))\,,
\pi_1\bd{\lambda}(t_0)\hat{\bd{g}}(t_0)(e^{\wedge\Phi}\wedge
\pi_1\bd{l}_2(e^{\wedge\Phi}))
\nonumber\\
&=\
\omega_l(\pi_1\bd{\lambda}(t_0)
(e^{\wedge g(\Phi)}\wedge g_{l_1}(e^{\wedge\Phi}))\,,
g_{l_2}(e^{\wedge\Phi}))
\nonumber\\
&\ \ \ 
+\omega_l(g_{l_1}(e^{\wedge\Phi})\,,
\pi_1\bd{\lambda}(t_0)(e^{g(\Phi)}\wedge g_{l_2}(e^{\wedge\Phi}))
\nonumber\\
&=\ 0\,,
\end{align}
where
\begin{equation}
 g(\Phi)\ =\ \pi_1\hat{\bd{g}}(t_0)(e^{\wedge\Phi})\,,\qquad
g_{l_i}(e^{\wedge\Phi})\ =\ \pi_1\hat{\bd{g}}(t_0)
(e^{\wedge\Phi}\wedge\pi_1\bd{l}_i(e^{\wedge\Phi}))\,,\quad
(i=1,2)\,.
\end{equation}
In the last equality we used the fact that
the gauge products $\bd{\lambda}$ represent the degree-even coderivation 
cyclic with respect to $\omega_l$\,:
\begin{equation}
 \omega_l(\Phi_1\,,\lambda_n(\Phi_2\wedge\cdots\wedge\Phi_{n+1}))\
=\ %(-1)^{\Phi_2+\cdots+\Phi_n+1}
(-1)^{|\Phi_1|}
\omega_l(\lambda_n(\Phi_1\wedge\cdots\wedge\Phi_n)\,,\Phi_{n+1})\,.
\end{equation}
Therefore the quantity
\begin{equation}
 \omega_l(\pi_1\hat{\bd{g}}(t_0)\bd{l}_1(e^{\wedge\Phi})\,,
\pi_1\hat{\bd{g}}(t_0)\bd{l}_2(e^{\wedge\Phi}))
\end{equation}
is independent of $t_0$\,, and in particular,
\begin{equation}
 \omega_l(\pi_1\hat{\bd{g}}(0)\bd{l}_1(e^{\wedge\Phi})\,,
\pi_1\hat{\bd{g}}(0)\bd{l}_2(e^{\wedge\Phi}))\ =\
 \omega_l(\pi_1\hat{\bd{g}}(1)\bd{l}_1(e^{\wedge\Phi})\,,
\pi_1\hat{\bd{g}}(1)\bd{l}_2(e^{\wedge\Phi}))\,. 
\end{equation}
%Taking into account
Since
$\hat{\bd{g}}(0)=\hat{\bd{g}}$ and $\hat{\bd{g}}(1)=\idSH$\,,
this is nothing but the identity in Eq.\,(\ref{id cyclic}).
%\begin{equation}
%\omega_l(\pi_1\hat{\bd{g}}\bd{l}_1(e^{\wedge\Phi})\,,
%\pi_1\hat{\bd{g}}\bd{l}_2(e^{\wedge\Phi}))\ 
%=\
%\omega_l(\pi_1\bd{l}_1(e^{\wedge\Phi})\,,
%\pi_1\bd{l}_2(e^{\wedge\Phi}))\,.
%\end{equation}

\end{document}